\documentclass[a4paper,11pt]{article}
\usepackage{jheppub} 
\usepackage{lineno}
\usepackage[normalem]{ulem} 
\usepackage{comment}
\usepackage{xcolor}
\usepackage{mathrsfs}
\usepackage{dsfont} 
\usepackage{tcolorbox} 
\usepackage{braket}
\usepackage{calligra}  
\usepackage{multirow}
\usepackage{soul}
\DeclareMathAlphabet{\mathcalligra}{T1}{calligra}{m}{n}

\newcommand*{\dt}{\mathrm{d}}
\newcommand*{\scri}{\ensuremath{\mathscr{I}}}

\usepackage{tikz}
\usepackage{amsmath} 
\usepackage{mathrsfs} 
\usepackage{xfp} 
\usepackage[outline]{contour} 

\title{Flat Holography \& Holographic Renormalization: Scalar Field}

\author{Martin Ammon,}
\author{Federico Capone,}
\author{Christoph Sieling}
\affiliation{Theoretisch-Physikalisches Institut, Friedrich-Schiller-Universität Jena,\\
Max-Wien-Platz 1, D-07743 Jena, Germany,}

\emailAdd{martin.ammon@uni-jena.de, federico.capone@uni-jena.de, christoph.sieling@uni-jena.de}

\abstract{We adapt the Hamilton-Jacobi method of holographic renormalization to scalar field theories in Minkowski spacetime with scattering boundary conditions. The approach yields a flat space holographic dictionary in which the expectation value of a dual operator is given by the renormalized canonical momentum. The source of the operator is imposed as a Dirichlet condition in a radial timelike foliation of the bulk theory and corresponds to the scattering data appearing in the Arefeva-Faddeev-Slavnov generating functional. We initiate a study of massive scalars and interacting fields within this formalism and we comment on extensions to different bulk theories and backgrounds.
}

\begin{document}
\maketitle
\flushbottom

\section{Introduction}

The AdS/CFT correspondence \cite{Maldacena:1997re} has deeply influenced our understanding of how the holographic principle is realised in a quantum theory of gravity with asymptotically AdS boundary conditions.  
The duality is dynamically defined by the Gubser-Klebanov-Polyakov-Witten prescription (GKPW) \cite{Gubser:1998bc, Witten:1998qj}, which equates the bulk partition function with the CFT generating functional once boundary data of the bulk theory are identified with sources for local gauge-invariant CFT operators. Banks, Douglas, Horowitz, and Martinec (BDHM) \cite{Banks:1998dd, Susskind:1998dq} introduced an alternative formulation of the AdS/CFT dictionary in which bulk QFT correlators are evaluated in a fixed AdS background and then extrapolated to the boundary with an appropriate rescaling. This so-called “extrapolate dictionary” has been shown to reproduce the GKPW prescription when applied carefully \cite{Harlow:2011ke}, even in the presence of static AdS black hole backgrounds \cite{Botta-Cantcheff:2019apr}.

The success of the AdS/CFT correspondence has long motivated the search for analogous holographic dualities for asymptotically flat quantum gravitational theories \cite{Polchinski:1999ry,Witten:2001kn,Arcioni:2003xx,deBoer:2003vf}. Such a duality is typically understood as a conjectured relation between two \emph{a priori} distinct theories -- for example, a string theory in the bulk and a non-string theory associated with the boundary -- whose dynamical matching can be tested in suitable regimes. However, in the absence of top-down constructions comparable to those underlying AdS/CFT, progress necessarily involves a degree of educated guesswork. Two guiding principles have emerged. First, the dual non-gravitational theory is expected to be strongly constrained by the bulk asymptotic symmetries, reflecting a deceptively simple rule from AdS/CFT. Second, an \emph{extrapolated dictionary} for flat holography can be constructed from effective scattering amplitudes. With appropriate integral transforms or limits of bulk fields to the boundary, the boundary symmetries are made manifest in the amplitude and the objects entering this transformed expression are taken to define the operators of the putative non-gravitational dual.

These arguments form the basis of the two leading approaches to flat holography today: the Celestial and Carrollian proposals. In the Celestial framework (see \cite{Pasterski:2021raf,McLoughlin:2022ljp, Donnay:2023mrd} for reviews), bulk scattering amplitudes in Minkowski spacetime are reformulated in a celestial basis using conformal primary wavefunctions, or equivalently via a Milne foliation of the bulk (e.g. \cite{Ball:2019atb,Melton:2023bjw,Iacobacci:2022yjo,Sleight:2023ojm}). This construction is applicable to any background admitting a homothetic Killing vector and has been extended to black hole spacetimes \cite{Gonzo:2022tjm}. Both massless and massive amplitudes can be expressed as Celestial correlators with slightly different transformations for massive versus massless states. 

The Carrollian approach (for reviews see \cite{Bagchi:2025vri,Nguyen:2025zhg}), instead, formulates amplitudes as correlators of a Conformal Carrollian theory on a Carrollian structure, typically identified with null infinity, $\mathscr{I}$. In this framework, massless states fit naturally, and the flat hologram is encoded in a codimension-one theory. However, massive scattering presents fundamental challenges, as no elementary representations of massive states at $\mathscr{I}$ are known  \cite{Have:2024dff} (see however \cite{Dappiaggi:2007hh} and \cite{Hao:2025btl}). Further insight into these proposals can be gained by examining flat limits of AdS/CFT, which offer a useful, albeit indirect, point of comparison (see e.g. \cite{Bagchi:2023fbj,Bagchi:2023cen,Campoleoni:2023fug,Alday:2024yyj,Lipstein:2025jfj,Fontanella:2025tbs}).

Recently, \cite{Kim:2023qbl,Kraus:2024gso} proposed that the  Arefeva, Faddeev, and Slavnov (AFS) generating functional  \cite{Arefeva:1974jv}  of massless scattering elements can serve as the basis for the construction of a Carrollian holographic dictionary in the spirit of the  GKPW rule (see also \cite{Jain:2023fxc} for a discussion  within the Celestial approach). The boundary conditions in the AFS functional are set by a free field at infinity, and in particular by its  positive frequency part (e.g. going as $\exp(-i\omega t)$ with $\omega=k^0>0$) fixed on a Cauchy slice in the far past and  its negative frequency part fixed on a Cauchy slice in the far future. The key insight of \cite{Kraus:2024gso,Jain:2023fxc} is to explicitly extrapolate these data to $\mathscr{I}^-$ and $\mathscr{I}^+$ using ingoing and outgoing coordinates, respectively. These then define the sources in a Carrollian generating functional, one for each insertion at  $\mathscr{I}^-$ and  $\mathscr{I}^+$.

While this approach is inspired by the GKPW dictionary, its basic mechanism sets it apart from it. This fact is nicely seen if we think about one of the defining features of AdS/CFT - elegantly captured by the GKPW dictionary - and how it might extend to flat holography. In AdS/CFT, the bulk ``emerges" from the boundary theory. This idea is encoded in the GKPW rule in the following way. For any given CFT state, the expectation value of any local (gauge invariant) operator in that state corresponds to the \emph{holographically renormalized} radial canonical momentum of the bulk theory \cite{Papadimitriou:2004ap,Papadimitriou:2007sj}. The renormalized momentum singles out one of the two asymptotic data of the bulk solutions. The other data corresponds to the source of the CFT operator. The CFT state in which the expectation value is evaluated corresponds, instead, to the conditions that relate the two asymptotic data in the interior of the bulk spacetime, which differ from spacetime to spacetime. Hence, the CFT state encodes information on the bulk spacetime. In contrast, the AFS path integral generally requires detailed knowledge of the bulk spacetime, as it is defined on the basis of a Cauchy slicing. 

The possibility of extending GKPW-type dictionaries beyond AdS/CFT, is closely linked to exploring how holographic renormalization can be generalized to other asymptotic geometries. In AdS/CFT, holographic renormalization renders the on-shell action finite and the variational problem well posed through the addition of local, covariant boundary counterterms. This procedure provides a systematic implementation of the GKPW prescription and  provides also  a framework for defining  conserved charges when the background subtraction method is not applicable \cite{Papadimitriou:2005ii,Hollands:2005wt,Hollands:2005ya}. 
These successes have motivated attempts to extend holographic renormalization to asymptotically flat spacetimes. However, due to the more intricate structure of the spacetime boundary, early works either employed hyperbolic foliations to adapt AdS techniques \cite{deBoer:2003vf,Solodukhin:2004gs,Costa:2012fm,Hao:2023wln} or focused solely on charges and actions at spacelike infinity \cite{Kraus:1999di,deHaro:2000wj,Mann:2005yr}, even suggesting a connection to little string theory in this context \cite{Marolf:2006bk}. Stimulated by the discovery of infrared triangles (e.g. \cite{Strominger:2017zoo} for an early review), more recent work has focused directly on null infinity and has revisited the notion of holographic renormalization in the context of both renormalizing the phase space at $\mathscr{I}^+$, where corner terms are crucial (e.g. \cite{Freidel:2019ohg,Chandrasekaran:2021vyu,Capone:2023roc,McNees:2023tus,Riello:2024uvs}), or directly the action \cite{Bagchi:2015wna,Hartong:2025jpp,Campoleoni:2025bhn}. However, to date, there exists no framework that holographically renormalizes bulk actions (or phase spaces) with an emergent radial direction and is, at the same time, directly connected to scattering boundary conditions. This limitation stems from the separate treatment of different boundaries (see also \cite{Compere:2023qoa}). 

\begin{figure}
\begin{center}
\usetikzlibrary{decorations.markings,decorations.pathmorphing}
\usetikzlibrary{angles,quotes} 
\usetikzlibrary{arrows.meta} 
\contourlength{1.4pt}

\newcommand{\calI}{\mathscr{I}} 
\tikzset{>=latex} 
\colorlet{myred}{red!80!black}
\colorlet{myblue}{blue!80!black}
\colorlet{mygreen}{green!80!black}
\colorlet{mydarkred}{red!50!black}
\colorlet{mydarkblue}{blue!50!black}
\colorlet{mylightblue}{mydarkblue!6}
\colorlet{mypurple}{blue!40!red!80!black}
\colorlet{mydarkpurple}{blue!40!red!50!black}
\colorlet{mylightpurple}{mydarkpurple!80!red!6}
\colorlet{myorange}{orange!30!yellow!29!white}

\tikzstyle{world line}=[myblue!60,line width=0.4] 

\tikzset{declare function={%
  penrose(\x,\c)  = {\fpeval{2/pi*atan( (sqrt((1+tan(\x)^2)^2+4*\c*\c*tan(\x)^2)-1-tan(\x)^2) /(2*\c*tan(\x)^2) )}};%
  penroseu(\x,\t) = {\fpeval{atan(\x+\t)/pi+atan(\x-\t)/pi}};%
  penrosev(\x,\t) = {\fpeval{atan(\x+\t)/pi-atan(\x-\t)/pi}};%
  kruskal(\x,\c)  = {\fpeval{asin( \c*sin(2*\x) )*2/pi}};
}}
\def\tick#1#2{\draw[thick] (#1) ++ (#2:0.04) --++ (#2-180:0.08)}
\def\Nsamples{20} 

\begin{tikzpicture}[scale=2]
  \message{Penrose diagram (radius r)^^J}
  
  \def\Nlines{4} 
  \def\ta{tan(90*1.0/(\Nlines+1))} 
  \def\tb{tan(90*2.0/(\Nlines+1))} 
  \coordinate (O) at ( 0, 0); 
  \coordinate (S) at ( 0,-1); 
  \coordinate (N) at ( 0, 1); 
  \coordinate (E) at ( 1, 0); 
  \coordinate (X) at ({penroseu(\tb,\tb)},{penrosev(\tb,\tb)});
  \coordinate (X0) at ({penroseu(\ta,-\tb)},{penrosev(\ta,-\tb)});
  
  \fill[myorange] (N) -- (E) -- (S) -- cycle;
  
  \node[left=6, right=6,black,align=center] at (1,0.04)
    { $i^0$};
  \node[above=2,below right=-2,black,align=left] at (0.04,-1)
    {$i^-$};
  \node[below=2,above right=-2,black,align=left] at (0.04,1)
    {$i^+$};
  \node[black,above right,align=right] at (57:0.68)
    {$\calI^+$};
  \node[black,below right,align=right] at (-60:0.68)
    {$\calI^-$};
  \foreach \i [evaluate={\c=\i/(\Nlines+1); \ct=tan(90*\c);}] in {2,...,\Nlines}{
    \message{  Running i/N=\i/\Nlines, c=\c, tan(90*\c)=\ct...^^J}
   
    \draw[world line,samples=\Nsamples,smooth,variable=\r,domain=-1:1] 
      plot({penrose(\r*pi/2,\ct)},\r);
  }
  \draw[thick,mydarkblue] (N) -- (E) -- (S) -- cycle;

  \draw[->,mydarkblue!80!black,shorten <=0.4] 
    ({penrose(-0.27*pi/2,tan(90*3/(\Nlines+1)))},-0.27) to[out=-55,in=170]++ (-35:0.3)
    node[right=-1] {$r=\text{constant}$};
 
\end{tikzpicture}
\end{center}
\label{fig:PenroseDiagram}
\caption{Penrose diagram of flat space with constant $r$ foliation. }
\end{figure}
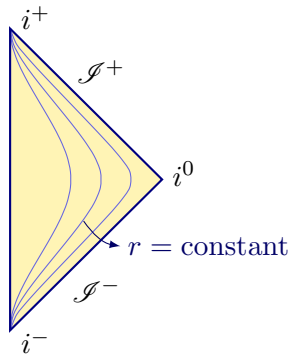

In this paper, we initiate a study of holographic renormalization in flat holography that departs significantly from existing approaches. Our goal is to use this method to construct a flat holography dictionary that remains faithful to the GKPW prescription, avoids the limitations of directly relying on the original AFS path integral, and refrains from using flat limits of the AdS/CFT correspondence. 

A central objective of our approach is to associate the putative dual field theory operator with a single bulk function that serves as its source in the ``boundary" path integral.  This single bulk function, corresponding to an appropriate term in the radial asymptotic expansion of the bulk field, encodes the scattering data $\phi_s$. Adapting the Hamilton-Jacobi holographic renormalization procedure 
(first proposed in \cite{deBoer:1999tgo,deBoer:2000cz}, further developed in \cite{Kalkkinen:2001vg,Martelli:2002sp} and further systematised and exemplified in \cite{Papadimitriou:2004ap,Papadimitriou:2007sj,Bergamin:2007sm,Papadimitriou:2010as,Elvang:2016tzz}) to this setting, we propose that the renormalized canonical momentum of the bulk theory corresponds to the expectation value of the dual operator in the presence of sources, in an appropriate sense to be specified later. Functional derivatives with respect to the source yield higher point correlators of the dual theory once bulk regularity conditions are imposed.

Our  scheme is similar to the logic of  GKPW and holographic renormalization, as it can be expressed as
\begin{align}
\label{eq:IntroDictionary}
    W_{\mathrm{CarrCFT}} [\phi_s] = S^{\mathrm{ren}}_{\mathrm{os}}[\phi_s]\, , 
\end{align}
where $ W_{\mathrm{CarrCFT}} $ is the generating functional of connected diagrams with sources $\phi_s$ and $S^{\mathrm{ren}}_{\mathrm{os}}$ is the bulk renormalized on-shell action evaluated as a functional of boundary data $\phi_s$. However, crucial technical differences persist, as we discuss in due time. 
The main point of this paper is to show that such a dictionary --  treating in some sense both past and future (null) infinities ``simultaneously" --  can indeed be achieved. Throughout sections \ref{sec:FreeScalarFieldSolutions}--\ref{sec:Holographicrenormalizationformasslessfields}, we exemplify the construction by considering a free, real scalar field in $(d+1)$-dimensional Minkowski spacetime in a radial timelike foliation defined by a coordinate $r$ (see figure \ref{fig:PenroseDiagram}), and comment on the extension of the framework to interacting fields, different backgrounds, and null foliations in sections \ref{sec:interactions} and \ref{sec:Discussion}.

To be specific, we exemplify most of our computations both in the standard $(t,r)$-coordinate system and in the Eddington–Finkelstein $(v,r)$-coordinates, promoting $r$ to be not only the coordinate used in asymptotic expansions, but also the direction of the Hamilton-Jacobi flow that determines the action. The purpose of this apparent redundancy of coordinates is to show that, although $(v,r)$-coordinates are most appropriate to study the asymptotic structure of fields at $\mathscr{I}^-$, the information on the asymptotic behavior at $\mathscr{I}^+$ is still encoded in a non-trivial manner 
on asymptotic $r=const$ slices, as indicated in section \ref{sec:FreeScalarFieldSolutions}. Although this observation may not appear surprising, it does not seem to have been widely utilized in the literature most closely associated with flat holography and infrared triangles. 
More importantly, when combined with the results of section \ref{sec:Holographicrenormalizationformasslessfields}, this observation leads us to a more abstract view of flat holography, one that is closer in spirit to AdS/CFT: the dual non-gravitational theory is defined on a (Carrollian) manifold that is not directly tied to any specific region of the spacetime asymptotic boundary. 

A brief outline of the main arguments, in the order they appear in the paper, is the following. 

In section \ref{sec:FreeScalarFieldSolutions}, we study the asymptotic solutions of the scalar field as $r\to\infty$. Starting in section \ref{sec:reviewAdS} with a brief review of the AdS case, we move to the Minkowskian case in section \ref{sec.FrobeniusThome}. The asymptotic solutions are obviously determined by two data, which we denote as $\phi^{\mathrm{(I)}}$ and $\phi^{\mathrm{(II)}}$. These are not the two independent coefficients of a Frobenius ansatz expansion as in the familiar context of AdS. In fact, $1/r=:{z}=0$ is an \emph{irregular singular point} of the radial equation associated to the Klein-Gordon equation, in contrast to the familiar AdS case. Hence, the asymptotic solutions take the form of Thomé expansions. These expansions are a power-law expansion times a term $\exp(\beta r)$, with $\beta$ depending on the frequency and the mass of the field, and on the coordinate choice. It is key to observe that in Lorentzian signature both branches of the solution possibly oscillate at infinity according to the frequency range in different coordinates. 

In section \ref{sec:generalHolographicRenormalization} we initiate our discussion of the action, opening in section \ref{sec:review_holoren} with the review of the scalar in AdS, where we indulge on the notion of locality of the counterterm. In section  \ref{sec:need_holo} we show that as a consequence of the Thomé solutions, the Lorentzian on-shell action is not well-defined in the asymptotic limit $r\to\infty$. The integrand defining the action strongly oscillates in this limit.  We impose an $i\epsilon$-prescription on the frequencies in accordance with the one used in Feynman propagators.  As a result, the branch of the solution associated to $\phi^{\mathrm{(II)}}$ diverges and the one associated to $\phi^{\mathrm{(I)}}$  is suppressed. With the chosen $i\epsilon$ prescription, e.g. a choice of boundary conditions, the on-shell action is divergent. In contrast to the familiar context of AdS/CFT, the divergences are \emph{exponential}. This implies crucial differences with respect to AdS/CFT holographic renormalization, as an infinite number of counterterms are needed. In other words,  the counterterm expansion does not truncate at finite order. It may then seem computationally hopeless to cure such divergences. Nonetheless, in section \ref{sec:HJ} we argue that the Hamilton-Jacobi approach, used in AdS/CFT as a neat organisation of the holographic renormalization procedure, remains a convenient tool for our purposes.

Section \ref{sec:Holographicrenormalizationformasslessfields} details the holographic renormalization procedure with the chosen examples. Sections \ref{sec:holoren_tr} and \ref{sec:holoren_vr} apply it  to the massless scalar field in our chosen coordinates.  Section \ref{sec:holographicCorrlators} presents the holographic two-point function with the necessary discussion. In particular, our result agrees with the existing extrapolated results in the literature, but our procedure selects a different relationship between the Carrollian source and the asymptotic value of the bulk field. This is not an accident and we comment on this apparent discrepancy in due course. 
We conclude with subsection \ref{sec:Massive} where we apply the Hamilton-Jacobi procedure to the massive scalar, showing that there is -- in some sense -- an imprint on $\scri$. 

Section \ref{sec:interactions} discusses an example of an interacting scalar field as a further extension of the formalism. In section \ref{sec:holo_ren_int} we comment on the additional divergences, once again drawing important lessons from AdS/CFT, especially from the case of irrelevant scalars. We do not carry out holographic renormalization. Instead, assuming that further divergences do not affect the free counterterm, section \ref{sec:propagators} lists the free propagators needed for the Witten–diagram computation introduced in section \ref{sec:3pt}. 

We conclude in section \ref{sec:Discussion} with a discussion of the results and further generalizations, together with additional comments situating our work within the existing literature where this was not addressed in the preceding sections. 

The paper contains six appendices.  Appendix \ref{app.ode} collects the main theorems concerning irregular singular points of ordinary differential equations, which supplement the material in section \ref{sec:FreeScalarFieldSolutions}. Appendix \ref{sec:RelationsAmongSolutions} expands on the various ways to write the scalar solution, which helps to elucidate the structure of the asymptotic data. Useful identities are collected in Appendix \ref{sec:Usefulidentities}. Appendix \ref{sec:EuclideanOnshellAction} is a complement to sections \ref{sec:generalHolographicRenormalization} and \ref{sec:Holographicrenormalizationformasslessfields}. It presents the scalar field action in Euclidean space, which is affected by genuine exponential divergences which are removed by the renormalization procedure. Appendix \ref{sec:factorization} provides a different way to construct the counterterms, relating them to a factorisation of the equations of motion. Appendix \ref{sec:InteractionDetails} displays steps omitted from section \ref{sec:interactions}.

\section{Free scalar field solutions}
\label{sec:FreeScalarFieldSolutions}
In this section we study the closed-form and asymptotic solutions of the free scalar equation
\begin{align}\label{eq:kgeq}
    \left( \Box - m^2 \right) \Phi =0
\end{align}
in Minkowski spacetime, drawing attention to the main differences compared to the field in AdS. From a strictly analytic perspective, the essential difference between the field equation in AdS spacetime and in Minkowski spacetime can be summarized as follows. When recast as a second-order ordinary differential equation in the radial coordinate, the boundary in AdS corresponds to a regular singular point, whereas in Minkowski spacetime it corresponds to an irregular singular point (see Appendix \ref{app.ode} for a brief review of the relevant ODE results). Consequently, in AdS the two independent asymptotic solutions appear as Frobenius series, possibly with logarithmic terms, whereas in Minkowski spacetime they are not simple Frobenius series. These solutions are known as Thomé solutions.  The connection to the plane wave solution is elaborated in Appendix \ref{sec:planewaveExpansion}.

\subsection{Review: the scalar field in AdS}\label{sec:reviewAdS}

 Since in the sections dedicated to Minkowski spacetime we will work with a radial coordinate $r$ such that asymptotic infinity is at $r\to\infty$, we will adopt the same convention here and write the AdS metric as
\begin{align}\label{eq:AdSmetric}
    \dt s^2  = \frac{\dt r^2}{r^2} + \gamma_{ab}\dt X^{a} \dt X^{b} =\frac{\dt r^2}{r^2} + r^2 \eta_{ab}\dt X^{a} \dt X ^{b}\, .
   \end{align}
In this coordinate system, the free massive scalar equation \eqref{eq:kgeq}
reads as
\begin{align}
  \left(r^2\partial_r^2+(d+1)r \partial_r + \frac{1}{r^2}\Box_{\eta}-m^2\right)\Phi=0 \, ,
\end{align}
and its asymptotic solutions around $r = \infty$ take the well-known form 
\begin{align}\label{eq.solutionAdS} 
\Phi(r,X)&=r^{-\Delta_-}\left(\phi_{(\Delta_-)}(X)+\dots\right) +r^{-\Delta_+}\left(\phi_{(\Delta_+)}(X)+\dots \right),
   \end{align}
   where
\begin{equation}\label{eq:DeltapmDef}
   	\Delta_\pm=\frac{d}{2}\pm\sqrt{\frac{d^2}{4}+m^2},\quad \Delta_-=d-\Delta_+,\quad \Delta_-\le\Delta_+\, . 
   \end{equation}
The ellipsis are subleading terms fully determined iteratively by the leading terms of the respective branch.  However, if the difference between $\Delta_+$ and $\Delta_-$ is an integer, the branch starting with $\phi_{(\Delta_+)}$ is not linearly independent from the other branch, as $\phi_{(\Delta_+)}$ is iteratively determined by $\phi_{(\Delta_-)}$. In such a case, the second branch must include logarithmic terms which restore the linear independence. The failure of linear independence is seen in the iterative procedure as a pole in the expression for $\phi_{(\Delta_+)}$. With the inclusion of the logarithmic term, the latter is determined by $\phi_{(\Delta_-)}$, while $\phi_{(\Delta_+)}$ remains undetermined. For the details of this procedure see, for example, \cite{Skenderis:2002wp}.

Although this story is well-known, we wish to emphasise that it ultimately stems from the fact that the point $r=\infty$ is a \emph{regular singular point} of the field equation,
as it is manifest 
using the definition and the theorem reported in Appendix \ref{app.ode}. Using the proper terminology for the asymptotic analysis of such equations, the two asymptotic expansions in AdS are of Frobenius type. 

The field equation can also be recognised to be a Bessel equation, which is hence (and with no surprise) exactly solvable in the form 
\begin{equation}
\label{eq:exactAdSsolution}
\Phi(r,X)=r^{-\frac{d}{2}}K_\nu\left(\frac{\sqrt{- \square_{\eta}}}{r}\right)\phi_1(X)+r^{-\frac{d}{2}}I_\nu\left(\frac{\sqrt{-\square_{\eta}}}{r}\right)\phi_2(X)\,,
\end{equation}
where $I_\nu$ and $K_\nu$ are the modified Bessel functions of the first and second kind, respectively, and 
$\nu= \sqrt{\frac{d^2}{4} + m^2}$. 
This solution is clearly formal. To unpack it, we would need to Fourier transform along the boundary directions so that $\Box_\eta$ becomes a momentum. The functions $K_\nu$ and $I_\nu$ behave asymptotically as the first and second branch of the asymptotic solution, respectively. Although related to the coefficients of the asymptotic solution, the terms of the expansion of $K_\nu$ and $I_\nu$ are not the same as the former. However, the closed-form solution reveals a further piece of information that is not available from the near boundary expansion: $I_\nu$ diverges in the interior and so regular solutions must have $\phi_2=0$. Crucially, this condition does not completely eliminate what we called second branch from the asymptotic solution, as the expansion of $K_\nu$ contains terms at those orders. Hence, at the level of the asymptotic solution, it implies that the interior boundary condition (regularity) imposes a \emph{non-local} relation between $\phi_{(\Delta_-)}$ and $\phi_{(\Delta_+)}$, which is a crucial ingredient of the holographic computation of CFT correlation functions.

In the AdS/CFT literature  the leading mode,  $\phi_{(\Delta_-)}$, is called non-normalizable and the independent subleading mode, $\phi_{(\Delta_+)}$, is called normalizable. In standard quantization, which requires usual Dirichlet conditions, the former plays the role of the source of a boundary operator whose vacuum expectation value is determined by the latter. In alternate quantization, with Neumann boundary conditions, the roles are swapped.

\subsection{Scalars in Minkowski spacetime}\label{sec.FrobeniusThome}
In section \ref{sec:Holographicrenormalizationformasslessfields} we will perform holographic renormalization for the scalar field in $(d+1)$-dimensional Minkowski space in two different coordinate systems, which we loosely refer to as $(t,r)$- and $(v,r)$-coordinates, with $v=t+r$, such that the metric reads 
\begin{align}
\label{eq:coordinates}
    \dt s^2&=-\dt t^2+ \dt r^2+ r^2 \dt \Omega^2_{d-1} =-\dt v^2+2\dt v \dt r+ r^2 \dt \Omega^2_{d-1}\,, 
\end{align}
where $\dt \Omega^2_{d-1} = \hat{g}_{ij}\dt \vartheta^i \dt \vartheta^j$ is the metric of the unit sphere $\mathds{S}^{d-1}$. Occasionally we will make comments about results in outgoing coordinates, which are obtained by $u = t-r$. The equations of motion for $\Phi(r,t,\Omega)$ and $\Phi(r,v,\Omega)$ read respectively as
\begin{equation}\label{eq.eom_tr}
    \ddot{\Phi} + \frac{d-1}{r} \, \dot{\Phi} + \left[\frac{1}{r^2} \Delta_{\mathds{S}^{d-1}} - \left( m^2+\frac{\partial^2}{\partial t^2} \right)\right] \Phi = 0 \, ,
\end{equation}
and 
\begin{align}\label{eq.eom_vr}
\ddot{\Phi}+\left(\frac{d-1}{r}+2\partial_v\right)\dot{\Phi}+\left[\frac{1}{r^2}\Delta_{\mathds{S}^{d-1}}-\left(m^2-\frac{d-1}{r}\partial_v\right)\right]\Phi =0\,, 
\end{align}
where the dot denotes the derivative with respect to $r$ in the following. These partial differential equations can be turned into a set of ordinary differential equations by Fourier transforming respectively in $t$ and $v$ and by decomposing in spherical harmonics on the sphere.  We use the following convention for the Fourier transform in $t$
\begin{equation}\label{eq.Fourier_convention}
\Phi(r, t,\Omega) = \frac{1}{2 \pi}\int_{-\infty}^{\infty} \, \dt \omega \, e^{-i \omega t} \,  \tilde{\Phi}(r,\omega,\Omega) 
\end{equation}
and similarly for the Fourier transform in $v$. For the spherical decomposition we use the real spherical harmonics $Y_{\ell, I}$ of the sphere $\mathds{S}^{d-1}$ satisfying $\Delta_{\mathds{S}^{d-1}} Y_{\ell, I} =-\ell(\ell+d-2) Y_{\ell,I}$ with $\ell \in \mathbb{N}_0.$ The index $I$ labels the different orthonormal spherical harmonics.\footnote{For $d=3$, i.e. in the case of the unit sphere $\mathds{S}^2$, the index $I$ is usually named $m$ taking values in $\{ -\ell, -\ell+1, \dots, \ell -1, \ell \}.$} Hence, we use
\begin{align}\label{eq:conventionsphericalharmonics}
    \tilde{\Phi}(r, \omega, \Omega) = \sum_{\ell, I}\,Y_{\ell, I}(\Omega) \, y_{\omega, \ell, I}(r)\,. 
\end{align}
Applying a Fourier transform as well as a spherical harmonics decomposition the differential equation in $(t,r)$-coordinates reads
\begin{equation}\label{eq.eomFourier_tr}
\left[\partial_r^2+\frac{d-1}{r}\partial_r-\left(\frac{1}{r^2}\ell(\ell+d-2)+m^2-\omega^2\right)\right]y_{\omega, \ell, I}=0\,. 
\end{equation}
Similarly, in $(v,r)$-coordinates using \eqref{eq.Fourier_convention} with $t$ replaced by $v$ and \eqref{eq:conventionsphericalharmonics} we obtain

\begin{equation}\label{eq.eomFourier_vr}
\left[\partial_{r}^2 +\left(\frac{d-1}{r} -2i\omega \right)\partial_r -\frac{1}{r^2}\ell(\ell+d-2) - \left(m^2  +\frac{d-1}{r}i\omega\right)\right]y_{\omega, \ell, I} =0\,. 
\end{equation}
For later discussion purposes, we will also consider the Euclidean space $\mathbb{R}^{d+1}$ with metric\footnote{Notice that in Euclidean signature it would appear natural to consider a metric of the form $\mathbb{R}_\rho \times \mathds{S}^d$, $\dt s^2=d\rho^2+\rho^2 \dt \Omega^2_d$, rather than the metric \eqref{eq:metrictaur} of the form $\mathbb{R}_\tau \times \mathbb{R}_r \times S^d$. The two options are clearly related by an obvious coordinate transformation, but the metric \eqref{eq:metrictaur} is more convenient for discussing analytical continuation to Lorentzian time.}
\begin{equation}\label{eq:metrictaur}
\dt s^2=\dt \tau^2 + \dt r^2 + r^2 \dt \Omega^2_{d-1}.
\end{equation}
In this case, the equation of motion for $\Phi(r,\tau,\Omega)$ is
\begin{equation}
\label{eq.eomtaur}
    \ddot{\Phi} + \frac{d-1}{r} \, \dot{\Phi} + \left[\frac{1}{r^2} \Delta_{\mathds{S}^{d-1}} - \left( m^2-\frac{\partial^2}{\partial \tau^2} \right) \right]\Phi = 0\,,
\end{equation}
which can be written as an ordinary differential equation (ODE) performing a Fourier transform in $\tau$ and a spherical-harmonics decomposition, using the same conventions with obvious substitutions. We get
\begin{equation}\label{eq.eomFourier_taur}
\left[\partial_r^2+\frac{d-1}{r}\partial_r-\left(\frac{1}{r^2}\ell(\ell+d-2)+m^2+\omega^2\right)\right]y_{\omega, \ell, I}=0.
\end{equation}
Notice (with no surprise, but it will be important) that the only difference between the Lorentzian $(t,r)$ equation and the Euclidean version is only a sign difference in front of the $\omega^2$ term.

We can now proceed to discuss the structure of the asymptotic solutions. 

\subsubsection{Asymptotic Thomé solutions}
\label{sec:SingularitiesOfDifferentialEquations}
The point $r = \infty$ is an irregular singular point\footnote{Note that for $\omega =0$, the point $r = \infty$ in the massless case is a regular singular point. } (for the definitions, see Appendix \ref{app.ode}) of the associated ODEs \eqref{eq.eomFourier_tr}, \eqref{eq.eomFourier_vr} and \eqref{eq.eomFourier_taur}. Accordingly, the asymptotic solutions are of Thomé type (Appendix \ref{app.solutionirreg}). Since the rank of the singular point is $2$, the two solutions can be written as
\begin{align}
\label{eq:AnsatzAsmyptoticExpansionGeneral}
    y_{\omega, \ell, I}(r) = e^{\beta r}\sum_{ k=0}^{\infty} \tilde{\phi}_{k}(\omega, \ell, I)\, r^{-k - \alpha}\,, 
\end{align}
with coefficients $\beta$, $\tilde{\phi}_{k}$  and $\alpha$ all depending on $(\omega, \ell, I)$ and differing for the two linearly independent solutions. 
The asymptotic expansion is determined iteratively and the coefficients depend on the spacetime signature, the coordinate system, and the mass of the field.

Instead of going to frequency space, we can formally also consider expansions of the form 
\begin{align}\label{eq:formalTR-ansatz}
\Phi = e^{\beta(\partial_{\tau}, \Delta_{\mathds{S}^{d-1}}) r}\sum_{k =0}^{\infty} c_{k}(\partial_{\tau}, \Delta_{\mathds{S}^{d-1}}) r^{-k-\alpha(\partial_{\tau}, \Delta_{\mathds{S}^{d-1}})}\phi(\tau, \Omega)\,. 
\end{align}
For $(t, r)$ and $(v, r)$-coordinates, the $\partial_{\tau}$ dependency is of course exchanged with a $\partial_t$ and $\partial_v$ dependence respectively. 
The operator valued functions $c_{k}, \beta$ and $\alpha$ can be obtained directly from the equations \eqref{eq.eom_tr}, \eqref{eq.eom_vr} and \eqref{eq.eomtaur}.  

Let us proceed by determining the coefficients $\beta, \alpha$ and $\tilde{\phi}_{k}$ in frequency space. 
We immediately see that $\beta$ is constrained by $m^2$ and $\omega$ in all cases. In fact, the derivatives of $y_{\omega, \ell, I}$ with respect to $r$ read 
\begin{equation}
\partial_r y_{\omega,\ell,I}=\sum_{k=0}^\infty \left(\beta+\frac{(-k-\alpha)}{r}\right)e^{\beta r}\tilde{\phi}_k r^{-k-\alpha}\,,
\end{equation}
\begin{equation}
\partial_r^2 y_{\omega,\ell,I}=\sum_{k=0}^\infty \left(\beta^2+\frac{2\beta(-k-\alpha)}{r}+\frac{(-k-\alpha)(-k-\alpha-1)}{r^2}\right)e^{\beta r}\tilde{\phi}_k r^{-k-\alpha}\,,
\end{equation}
which, once substituted in the equations of motion (see equations \eqref{eq.eomFourier_taur}, \eqref{eq.eomFourier_tr} and \eqref{eq.eomFourier_vr} for $(\tau, r)$, $(t, r)$ and $(v,r)$-coordinates), give at order $k=0$ (generic $\alpha$)
the constraints 
\begin{subequations}
\begin{align}
&\text{ Euclidean } (\tau,r): &&\beta^2-(m^2+\omega^2)=0\,,\label{eq:beta_Euclidean}\\
& \text{ Lorentzian } (t,r): &&\beta^2-(m^2-\omega^2)=0 \,, \label{eq:beta_tr}\\
& \text{ Lorentzian } (v,r):&&\beta^2-2i\omega \beta-m^2=0\,. \label{eq:beta_vr}
\end{align}
\end{subequations}
Only for visualization purposes we list first the Euclidean $(\tau,r)$-coordinates followed by the Lorentzian $(t,r)$ and $(v,r)$-coordinates. 
Once this is determined, the order $k=0$ fixes $\alpha$ in all cases to be
\begin{equation}
\alpha=\frac{d-1}{2}\,.\label{eq:DefAlpha}
\end{equation}
The coefficients $\tilde{\phi}_k$ are related by a recursion relation. For both the Euclidean $(\tau,r)$ and the Lorentzian $(t,r)$ equations we have
\begin{align}
\label{eq:EucldieanRecursionRelation}
    \tilde{\phi}_{k+1} = \frac{(\alpha + k)(\alpha + k+2-d) -\ell(\ell +d -2)}{2( k+1)\beta}\tilde{\phi}_k\,  , 
\end{align}
where $\beta$ is to be chosen as the solution of \eqref{eq:beta_Euclidean} or \eqref{eq:beta_tr}, accordingly. For the field equations written in $(v,r)$-coordinates we have
\begin{equation}
\label{eq:EucldieanRecursionRelationvr}
\tilde{\phi}_{k+1}= \frac{(\alpha +k)(\alpha + k + 2 - d) - \ell(\ell +d-2)}{2( \beta - i \omega)(k+1)}\tilde{\phi}_k\,. 
\end{equation}
Since the equation of motion is of second order, there are two linearly independent solutions, which we call branches $(\mathrm{I})$ and $(\mathrm{II})$. The recursion relations do not fix $\tilde{\phi}_{0}(\omega,\ell,I)$, which is the free data in each branch. We denote them respectively $\tilde{\phi}^{(\mathrm{I})}(\omega,\ell,I)$ and $\tilde{\phi}^{(\mathrm{II})}(\omega,\ell,I)$ dropping the index. Each branch is also associated with a solution of the equations \eqref{eq:beta_Euclidean}, \eqref{eq:beta_tr} and \eqref{eq:beta_vr} which we denote as $\beta_\pm$. The expansion of the field in $(t,r)$-coordinates thus reads as\footnote{Note that \eqref{eq:TRasymptoticFullField} does not imply that $\Phi(r, t, \Omega) \sim r^{-\frac{d-1}{2}} \left(\phi(t, \Omega) + \mathcal{O}(r^{-1})\right)$, see also the formal solution \eqref{eq:TRasymptoticFullFieldRepackaged}.  }
\begin{align}
\label{eq:TRasymptoticFullField}
\Phi(r,t,\Omega)&=\frac{1}{2\pi}\int\limits_{-\infty}^{+\infty}\dt\omega e^{-i\omega t} \tilde{\Phi}(r,\omega,\Omega)=\frac{1}{2\pi}\int\limits_{-\infty}^{+\infty}\dt \omega e^{-i\omega t}\sum_{\ell,I}Y_{\ell,I}\,y_{\omega,\ell,I}\nonumber\\
&=\frac{1}{2\pi}\int\limits_{-\infty}^{+\infty}\dt\omega e^{-i\omega t}\sum_{\ell,I}Y_{\ell,I}\left[\frac{e^{\beta_+r}}{r^{\frac{d-1}{2}}}\left(\tilde{\phi}^{(\mathrm{I})}(\omega, \ell, I)+\dots\right)+\frac{e^{\beta_-r}}{r^{\frac{d-1}{2}}}\left(\tilde{\phi}^{(\mathrm{II})}(\omega, \ell , I)+\dots\right)\right]\, .
\end{align}
The same expression also applies in the Euclidean $(\tau,r)$-coordinates and in the $(v,r)$-coordinates, provided one performs the formal replacements $t \to \tau$ and $t \to v$, respectively, together with the corresponding choice of the coefficients $\beta$ appropriate to each case. Explicitly, the solutions of \eqref{eq:beta_Euclidean}, \eqref{eq:beta_tr} and  \eqref{eq:beta_vr} are
\begin{subequations}
\begin{align}
&\text{Euclidean } (\tau,r): &&  \beta_{\pm}=\pm\sqrt{m^2+\omega^2},\\  &\text{Lorentzian } (t,r): && \beta_{\pm}=\pm i \sqrt{\omega^2-m^2} \label{eq:betamassive_tr},\\
&\text{Lorentzian } (v,r): && \beta_{\pm}=i\omega\pm i\sqrt{\omega^2-m^2}\,,\label{eq:BetaMassiveVR}
\end{align}
\end{subequations}
where the choice of $\beta$ in the Lorentzian cases stem from the fact that solutions which are regular in the interior have $|\omega|\ge m$, as elaborated in Appendix \ref{sec:planewaveExpansion}. In the massless limit, the above relations for $\beta_{\pm}$ become\footnote{Note that solving directly the massless wave equation we would infer $\beta_+ = \omega$ and $\beta_- = - \omega$ in case of $(\tau, r)$- coordinates, or $\beta_+=2i\omega$ and $\beta_-=0$ in the Lorentzian $(v,r)$ case. 
As will become clear in the remainder of this section, it will be convenient to work with \eqref{eq:betamassless_euc}, \eqref{eq:betamassless_tr}, \eqref{eq:betamassless_vr} also for massless fields. }
\begin{subequations}
\begin{align}
   &\text{Euclidean } (\tau,r): &&  \beta_{\pm}=\pm |\omega|, \label{eq:betamassless_euc} \\
   &\text{Lorentzian } (t,r): && \beta_{\pm}=\pm i |\omega| , \label{eq:betamassless_tr} \\
   &\text{Lorentzian } (v,r): && \beta_{\pm}=i\omega\pm i|\omega| \label{eq:betamassless_vr}
\end{align}
\end{subequations}
 The asymptotic behaviour of the field as $r \to \infty$ is therefore as follows: it is exponentially divergent in the Euclidean signature, oscillatory in the Lorentzian $(t,r)$-coordinates, and in the advanced $(v,r)$-coordinates it becomes either oscillatory or decays as a power law depending on the sign of the frequency. Specifically, the $\mathrm{(I)}$-branch corresponds to $\beta_+ = 0$ for $\omega < 0$ and is therefore non-oscillatory, while for $\omega > 0$ it is the $\mathrm{(II)}$-branch that becomes non-oscillatory. This behaviour is summarized in table \ref{tab:FrequenciesBranches}, in parallel with its map to retarded, $(u,r)$-coordinates. Notice that the branches are interchanged. For positive frequencies in $(u,r)$-coordinates, the $\mathrm{(I)}$-branch comes with $\beta_+ =0$ and is hence not oscillating, while for negative frequencies, the $\mathrm{(II)}$-branch is not oscillating. The scattering data are the positive frequency part at $\scri^-$ and the negative frequency part at $\scri^+$. They correspond to the non-oscillating part of the solution in the coordinate systems adapted to $\scri^-$ and $\scri^+$, respectively $(v,r)$- and $(u,r)$-coordinates. In order to capture scattering data while working within a single coordinate system, it is hence not feasible to restrict only to Frobenius-type solutions; rather we need the Thomé solutions.  The on-shell action is greatly affected by this, as detailed in \ref{sec:need_holo}, and this is the fundamental reason for the development of the rest of this paper. 
 
\begin{table}[t]
        \centering
        \setlength{\tabcolsep}{20pt}
\renewcommand{\arraystretch}{2.5}
\begin{tabular}{ c|c c|c}
           $(v,r)$-coordinates &$\omega < 0$ &$\omega > 0$ &$(u,r)$-coordinates  \\  \hline
             (I)-branch& $\dfrac{1}{r^{\frac{d-1}{2}}}$ & $\dfrac{e^{\pm 2i\omega r}}{r^{\frac{d-1}{2}}}$ &    (II)-branch \\    
           (II)-branch & $\dfrac{e^{\pm 2i\omega r}}{r^{\frac{d-1}{2}}}$& $\dfrac{1}{r^{\frac{d-1}{2}}}$   &(I)-branch      
        \end{tabular}
        \caption{Leading behavior of the two asymptotic branches of free scalar field in the large $r$ limit in retarded and advanced coordinates. The upper sign corresponds to $(v, r)$-coordinates, while the lower sign corresponds to $(u,r)$-coordinates. \label{tab:FrequenciesBranches}} 
\end{table}

It may sometimes be convenient to use the more formal expression introduced above in \eqref{eq:formalTR-ansatz}. The differential operators $\beta_{\pm}$ in the different coordinates/signatures read
\begin{subequations}
\begin{align}
   &\text{Euclidean } (\tau,r): &&  \beta_{\pm}=\pm \sqrt{ - \partial_{\tau}^2 +m^2}\,, \label{eq:betamassive_eucFORMAL}\\  &\text{Lorentzian } (t,r): && \beta_{\pm}=\pm i \sqrt{-\partial_t^2 - m^2} \,, \label{eq:betamassive_trFORMAL}\\
&\text{Lorentzian } (v,r): && \beta_{\pm}=- \partial_v\pm i\sqrt{-\partial_v^2 - m^2} \, .\label{eq:betamassive_vrFORMAL}
\end{align}
\end{subequations}
This allows a repackaging of the field expansion \eqref{eq:TRasymptoticFullField} as
\begin{equation}\label{eq:TRasymptoticFullFieldRepackaged}
\Phi(r,t,\Omega)=\frac{1}{r^{\frac{d-1}{2}}}\left[e^{\beta_+ r}\left(\phi^{(\mathrm{I})}(t,\Omega)+\dots\right)+e^{\beta_- r}\left(\phi^{(\mathrm{II})}(t,\Omega)+\dots\right)\right],
\end{equation}
where $\phi^{(\mathrm{I})}$ and $\phi^{(\mathrm{II})}$ represent the two independent modes of the asymptotic expansion. 

\subsubsection{Closed form solutions}
In this section we solve the differential equations for all values of $r$, and not just asymptotically at $r = \infty$ as done before. We first focus on $(t,r)$-coordinates. In fact, the solution to equation \eqref{eq.eomFourier_tr} is given in terms of the Hankel functions\footnote{Alternatively, we can express the solution in terms of the Bessel function of the first kind, $J_{\frac{d-2}{2} + \ell}( r\sqrt{\omega^2-m^2})$, and of the Bessel function of the second kind, $Y_{\frac{d-2}{2} + \ell}( r\sqrt{\omega^2 - m^2})$. However, as we will see, the Hankel functions are easier to deal with when it comes to asymptotic expansions.} of the first and second kind
\begin{align}\label{eq:solutiontr}
y_{\omega, \ell, I}(r) = r^{-\frac{d-2}{2}} \, \left( \tilde{\phi}_{1}(\omega, \ell, I) \, H^{(1)}_{\frac{d-2}{2} + \ell}(r\sqrt{\omega^2 - m^2}) + \tilde{\phi}_{2}(\omega, \ell, I)\,  H^{(2)}_{\frac{d-2}{2}+\ell}(r \sqrt{\omega^2 - m^2})\right)
\end{align}
with $\tilde{\phi}_{1}(\omega, \ell, I)$ and $\tilde{\phi}_{2}(\omega, \ell, I)$ being complex numbers, that are related to each other once we discuss regularity. The general solution $\Phi(r,t,\Omega)$ is obtained by inserting \eqref{eq:solutiontr} into \eqref{eq:conventionsphericalharmonics} and \eqref{eq.Fourier_convention}
\begin{align}\label{eq:solutiongeneraltr}
\begin{split}
\Phi(r, t,\Omega) = r^{-\frac{d-2}{2}}\sum_{\ell, I} \, \int\limits_{-\infty}^{\infty} \, \frac{\dt \omega}{2\pi} \, &\left( \tilde{\phi}_{1}(\omega, \ell, I) \, H^{(1)}_{\frac{d-2}{2} + \ell}(r\sqrt{\omega^2 - m^2}) \right. \\ &\quad \left. +\,  \tilde{\phi}_{2}(\omega, \ell, I) \,  H^{(2)}_{\frac{d-2}{2}+\ell}(r \sqrt{\omega^2 - m^2})\right) \, e^{-i \omega t} \,Y_{\ell, I}(\Omega) \,.
\end{split}
\end{align}
We can rewrite the general solution in a more formal way as 
\begin{align}
\label{eq:Formal_massive_tr_solution}
    \Phi(r,t,\Omega) = r^{-\frac{d-2}{2}} \, \left(H^{(1)}_{\nu}\left(\sqrt{-\partial_t^2 - m^2} \, r\right) \, \phi_1(t,\Omega) + H^{(2)}_{\nu}\left(\sqrt{-\partial_t^2 - m^2} \, r\right) \, \phi_{2}(t,\Omega)\right)\, ,  
    \end{align}
where $\nu$ is given by
\begin{equation}\label{eq:nuDef}
\nu = \sqrt{\frac{(d-2)^2}{4} - \Delta_{\mathds{S}^{d-1}}} \,.
\end{equation}
Note that in the formal solution \eqref{eq:Formal_massive_tr_solution} both the order $\nu$ of the Hankel function and its argument are differential operators. The relationship between the coefficients $\tilde{\phi}_{1}(\omega, \ell, I)$ appearing in \eqref{eq:solutiongeneraltr} and $\phi_1(t,\Omega) $ in \eqref{eq:Formal_massive_tr_solution} (similarly between $\phi_2(t,\Omega)$ and $\tilde{\phi}_{2}(\omega, \ell, I)$) is 
\begin{equation}
\phi_1(t,\Omega) =   \sum_{\ell, I} \, \int\limits_{-\infty}^\infty \frac{\mathrm{d} \omega}{2\pi} \, e^{-i \omega t} \, \tilde{\phi}_{1}(\omega, \ell, I) \,Y_{\ell, I}(\Omega)\,.
\end{equation}
The solutions given above are in general not regular at $r=0$. Regularity imposes that   $\tilde{\phi}_{1}( \omega, \ell, I) = \tilde{\phi}_{2}( \omega,\ell, I )$ for all $\omega, \ell$ and $I$, or, in terms of the functions  $\phi_1(t,\Omega)$ and $\phi_2(t,\Omega)$, the equality
\begin{align}
\label{eq:regularityInterior}
 \phi_1(t,\Omega) =\phi_2(t,\Omega)\,.
\end{align}
In fact, only the particular combination $H^{(1)}_{\nu}(\beta r) + H^{(2)}_{\nu}(\beta r) =2 J_{\nu}(\beta r)$ of the two Hankel functions $H^{(1)}_{\nu}(\beta r)$ and $H^{(2)}_{\nu}(\beta r)$ is regular at $r=0$. 
Thus, in contrast to the asymptotic solution, the regular solution of the scalar field reads 
\begin{align}
    \label{eq:regularsolwithsource}
    \Phi(r, t, \Omega)  = r^{-\frac{d-2}{2}}\, \sum_{\ell, I}\,\int\limits_{|\omega| \geq m} \frac{\dt \omega}{\pi}\, e^{- i \omega t}\, J_{\nu}(\sqrt{\omega^2 - m^2}\,r) Y_{\ell, I}(\Omega) \,\tilde{\phi}_2(\omega, \ell, I)\, , 
\end{align}
which can also be obtained from the plane wave expansion \eqref{eq:modedecompBessel}. 
The $\nu$ appearing in \eqref{eq:regularsolwithsource} is the eigenvalue of the operator \eqref{eq:nuDef}, which is given by
\begin{align}
\label{eq:EigenvalueNU}
    \nu = \ell + \frac{d-2}{2}\, . 
\end{align}

The corresponding closed-form solutions in $(v,r)$-coordinates are obtained by replacing $t$ with $v-r$
\begin{align}\label{eq:solutiongeneralvr}
\begin{split}
\Phi(r, v,\Omega) = r^{-\frac{d-2}{2}} \sum_{\ell, I} \, \int\limits_{-\infty}^{\infty} \, \frac{\dt \omega}{2\pi} \, \left( \tilde{\phi}_{1}(\omega, \ell, I) \, H^{(1)}_{\frac{d-2}{2} + \ell}(r\sqrt{\omega^2 - m^2})\right.\\\qquad +\left. \tilde{\phi}_{2}(\omega, \ell, I) \,  H^{(2)}_{\frac{d-2}{2}+\ell}(r \sqrt{\omega^2 - m^2})\right) \,Y_{\ell, I}(\Omega) \,e^{-i\omega v + i \omega r}\,,
\end{split}
\end{align}
or in a more formal way 
\begin{align}
\label{eq:Formal_massive_vr_solution}
    \Phi(r,v,\Omega) = r^{-\frac{d-2}{2}} \, e^{-\partial_v r}\left(H^{(1)}_{\nu}(\sqrt{-\partial_v^2 - m^2} r) \, \phi_1(v,\Omega) + H^{(2)}_{\nu}(\sqrt{-\partial_v^2 - m^2} r) \, \phi_{2}(v,\Omega)\right), 
\end{align}
where $\nu$ is still given by \eqref{eq:nuDef}.

\section{The on-shell action: general considerations}
\label{sec:generalHolographicRenormalization}
We now delve into the task of evaluating the action for the free scalar field. We introduce the notation 
\begin{equation}\label{eq:generalscalarfieldaction}
S=\int_{\mathds{M}_{d+1}}\dt^{d+1}x\,L=\int_{\mathds{M}_{d+1}}\dt^{d+1}x\sqrt{|g|}\mathscr{L},\qquad \mathscr{L}=\frac{\mathtt{s}}{2}(g^{\mu \nu} \partial_{\mu} \Phi \partial_{\nu}\Phi-m^2 \Phi^2)
\end{equation}
where $\mathtt{s}=\pm1$ respectively for Euclidean or  Lorentzian signature. On shell, it reduces to a boundary term
\begin{equation}
\label{eq:generalscalarfieldOnshellAction}
 S_{\text{os}}=  \frac{\mathtt{s}}{2}\int_{\mathds{M}_{d+1}} \dt^{d+1}x\,\partial_{\mu}\left( \sqrt{|g|}\, g^{\mu \nu} \, \Phi \partial_{\nu}\Phi\right)\,. 
\end{equation}
Any of the coordinate systems considered in the previous section foliate the spacetime with surfaces of constant $r$, so they can all be cast in ADM-like form
\begin{equation}\label{eq:foliation}
\mathrm{d}s^{2}=N^{2}\,\mathrm{d}r^{2}+h_{ab}\bigl(\mathrm{d}X^{a}+N^{a}\mathrm{d}r\bigr)\bigl(\mathrm{d}X^{b}+N^{b}\mathrm{d}r\bigr),
\end{equation}
with appropriate choices of the lapse, $N$, the shift, $N^{a}$, and the induced metric on the slices with coordinates $X^{a}$, $h_{ab}$. The boundary of interest is given by a surface $r=r_0$.\footnote{We will neglect any future or past boundary term, or corners.} The regulated on-shell action reads 
\begin{align}
\label{eq:def:regOS}
    S_{\mathrm{os}}^{\mathrm{reg}} : = \frac{\mathtt{s}}{2}\int_{r=r_0}\dt^d X \sqrt{|h|} \Phi \partial_n\Phi\,,  
\end{align}
where $\partial_n=n\cdot\partial$ is the derivative along the unit normal, $n$, to $r=r_0$. 

In section \ref{sec:review_holoren} we briefly review holographic renormalization in AdS, highlighting the main points to keep in mind when we extend the formalism to flat spacetime. Then, in section \ref{sec:need_holo} we discuss \emph{why} a procedure of holographic renormalization is generically needed in (asymptotically) flat spacetime by showing that the scalar on-shell action is ill-defined in an infinite volume of spacetime. We further track crucial elements differentiating the flat problem from the AdS problem. In section \ref{sec:HJ} we explain how we use the Hamilton-Jacobi formalism for the purpose of obtaining appropriate counterterms and the final renormalised action. While the root of the method is evident and its usage in AdS/CFT is well-known \cite{deBoer:1999tgo,Papadimitriou:2004ap,deBoer:2000cz, Papadimitriou:2007sj}, the way we employ it in Lorentzian flat space is entirely novel. The details of the specific examples chosen are deferred to section \ref{sec:Holographicrenormalizationformasslessfields}.

\subsection{Review: Holographic renormalization in AdS and the AdS/CFT dictionary}\label{sec:review_holoren}
In the given coordinates \eqref{eq:AdSmetric},  the on-shell action of a free massive scalar field in Euclidean AdS, 
\small
\begin{equation}
\label{eq:AdSbareOnShellaction}
 \begin{split}
 S^{\mathrm{reg}}_{\text{os}}=\int_{r=r_0} \dt^d X  \sqrt{|\eta|}\,r^{2\Delta_+ -d}\left( \frac{(\Delta_+-d)}{2} \phi_{(d-\Delta_+)}^2 - \frac{1}{r^2}\frac{d-\Delta_+ +1 }{2(2 \Delta_{+} - d-2)}\phi_{(d-\Delta_+)} \square_{\eta} \,\phi_{(d-\Delta_+)}+\dots\right),  
\end{split}
\end{equation}
\normalsize
manifestly diverges as $r_0\to+\infty$. Holographic renormalization modifies the action with the addition of local and covariant boundary counterterms $S^{\mathrm{ct}}$ that make the new action 
\begin{align}
\label{eq:generalCounterterms}
    S^{\mathrm{ren}}[\phi_{(d-\Delta_+)}] := \lim_{r_0 \to \infty} \, \left(S^{\mathrm{reg}}_{\mathrm{os}}\left[ \Phi(r_0), r_0\right] - S^{\mathrm{ct}}\left[ \Phi(r_0),r_0\right]\right)\,,
\end{align}
finite on-shell. The combination in parenthesis, before taking the limit, is usually called subtracted action $S^{\mathrm{sub}}[\Phi,r_0]$. The two keywords of the process are 
\begin{itemize}
    \item Locality: The counterterm should only involve the field and a finite number of longitudinal derivatives along the surface $r=r_0$.
    \item Covariance: The counterterm should be invariant under diffeomorphisms. 
\end{itemize}
The only possible counterterm action that satisfies these requirements is a finite polynomial of the \emph{induced} field on $r=r_0$, $\Phi(r_0,X)$, and its $\Box_\gamma$-derivatives. This differentiates this procedure from a naive subtraction of divergent orders. Specifically, the counterterm can be written as
\begin{equation}
\label{eq:AdSAnsatzCountertermAction}
S^{\text{ct}}=\frac{1}{2}\int_{r=r_0} \dt^d X \sqrt{|\gamma|}\,\Phi f(\Box_\gamma)\Phi\,,
\end{equation}
where $f$ is a polynomial. For example, when the operator has dimension $\frac{d}{2} + 1 < \Delta < \frac{d}{2} +2$, the only divergences are the two shown in \eqref{eq:AdSbareOnShellaction} and the appropriate $f$ is\footnote{Depending on $\Delta$, $f$ may include higher powers of $\Box_\gamma$. Terms like $(-\Box_\gamma)^n \log(-\Box_\gamma)$ appear when $\Delta = \frac{d}{2} + n$ with $n \in \mathbb{N}$; these  terms can be split into a local part and a finite order nonlocal piece in $\Box_\gamma$, which is then omitted from the counterterms. }
\begin{equation}\label{eq:example_f}
f(\Box_\gamma) =-(d-\Delta_+)-\frac{\Box_\gamma}{(2\Delta_+-d-2)}\,.
\end{equation}
Among the various ways to determine $f$, perhaps the most insightful is to recast the problem as an asymptotic Hamilton-Jacobi equation with the flow along the holographic direction $r$. It is important to stress that in order to preserve covariance formally, the $r$ dependence has been traded for the covariant object $\Box_\gamma$. For details see \cite{Papadimitriou:2010as}.  Another observation to keep in mind is that the function $f(\Box_\gamma)$ gives rise, in general, to a non-local counterterm in \eqref{eq:AdSAnsatzCountertermAction}. Only when truncating its expansion in powers of $\Box_\gamma$ (namely $r$), as for example in \eqref{eq:example_f}, we get a manifestly local counterterm.  

Once the regularisation procedure has been carried out, the renormalised on-shell action 
in our specific example reads as
\begin{equation}\label{eq:AdS_onshell_partially}
S^{\mathrm{sub}}[\Phi]=\int \dt^d X\sqrt{|\gamma|}\,\Phi\,(\partial_r-f)\,\Phi\,,\qquad S^{\mathrm{ren}}[\phi_{(d-\Delta_+)}]=\mathcal{N}\int \dt^d X\sqrt{|\eta|}\,\phi_{(d-\Delta_+)}\phi_{(\Delta_+)}\,.
\end{equation}
We have deliberately introduced the symbol $\mathcal{N}$ for the apparently innocuous numerical factor $\mathcal{N} = 2 \Delta_+-d$, to emphasize that a naive subtraction of divergences would yield a different value, and an incorrect value of $\mathcal{N}$ would lead to violations of the dual CFT Ward identities. The importance of such coefficients was first noted in \cite{Freedman:1998tz} and revisited in \cite{Klebanov:1999tb}, where appropriate boundary terms were added to the action to get the right result. Holographic renormalization automatically accounts for this.

One more thing to stress is that \eqref{eq:AdS_onshell_partially} is only formally on-shell at this point. No assumption on the bulk behaviour of the field has been made yet. Writing $S^{\mathrm{ren}}[\phi_{(d-\Delta_+)}]$ is, at this level, an abuse of notation because the asymptotic analysis does not relate $\phi_{(\Delta_+)}$ and $\phi_{(d-\Delta_+)}$. Once regularity in the interior is imposed, one obtains a non-local relation between $\phi_{(\Delta_+)}$ and $\phi_{(d-\Delta_+)}$ and hence the actual on-shell action is a \emph{non-local} functional of the source $\phi_{(d-\Delta_+)}$.

The asymptotic form of the renormalized on-shell action, e.g. \eqref{eq:AdS_onshell_partially} without any bulk condition, can be neatly interpreted  as 
\begin{equation}
S^\mathrm{sub}=\int \dt^d X\,\Pi^{\mathrm{sub}}\Phi\, , 
\end{equation}
where we have defined 
\begin{equation}
\label{eq:AdScanonicalMomentum}
\Pi^{\mathrm{sub}}=  \sqrt{|\gamma|}(\partial_n-f)\Phi \,.
\end{equation}
This is the subtracted momentum, namely the radial momentum
\begin{equation}
\Pi=\frac{\partial L}{\partial\dot{\Phi}}=\sqrt{|\gamma|}\,\partial_n\Phi\,,
\end{equation}
from which we subtract $\sqrt{|\gamma|}f\Phi=\frac{\delta S^{\mathrm{ct}}}{\delta \Phi}$. We come back to this in subsection \ref{sec:HJ} as this is the key to the Hamilton-Jacobi holographic renormalization strategy. The one-point function in the presence of sources is then defined as
\begin{equation}\label{eq:AdSvev}
\langle \mathcal{O}\rangle_{\phi_{(d-\Delta_+)}}  =  \lim_{r_0 \to \infty} r_0^{\Delta_+}\Pi^{\mathrm{sub}} = \mathcal{N} \phi_{(\Delta_+)}\,,  
\end{equation}
stressing that no condition on the behaviour of the bulk field in the interiror of the spacetime is required \cite{Papadimitriou:2004ap}. Higher-point functions are obtained by taking further derivatives of the canonical momentum with respect to the source. This time the bulk regularity conditions are strictly needed and hence,  a two-point function is indeed the second variational derivative of the on-shell action.

Let us conclude this review with a small comment concerning interacting scalars in the bulk. In order to find the bulk solution, one treats the coupling $\lambda$ perturbatively and solves order by order in $\lambda$ perturbing the free-field solution. If the bulk scalar is dual to a relevant CFT operator ($\Delta<d$), no new bulk divergences arise beyond those of the free field action. In contrast, for irrelevant CFT operators ($\Delta>d$), the bulk interacting scalar action exhibits novel divergences, which are treated with \emph{(pseudo)local} counterterms \cite{vanRees:2011fr,vanRees:2011ir}. A similar story holds for some marginal operators \cite{Castro:2024cmf}.

\subsection{The case for holographic renormalization in flat spacetime: why and how}\label{sec:need_holo}
Let us now consider the scalar field in Minkowski spacetime. 
In $(t,r)$- and $(\tau,r)$-coordinates, using the solution \eqref{eq:TRasymptoticFullFieldRepackaged}, the on-shell action takes the form\footnote{The formal multiplication is defined as
$
\partial_t f(t) \cdot \partial_t g(t) =  (\partial_t f(t)) (\partial_{t'} g(t'))_{| t'=t} 
$. In other words, the differential operator acts only on the function immediately to its right.} 
\begin{align}\label{eq:reg_action_formal}
\begin{split}
 S^{\mathrm{reg}}_{\mathrm{os}} = \frac{\texttt{s}}{2}\int \dt^d X \sqrt{\hat{g}} \,  &\left[e^{\beta_+ r_0}\phi^{(\mathrm{I})} \cdot e^{\beta_+ r_0} \left(\beta_+ +O(r_0^{-1})\right) \phi^{(\mathrm{I})} \right. \\ & \left. +e^{\beta_- r_0}\phi^{(\mathrm{II})} \cdot e^{\beta_- r_0}\left(\beta_-+O(r_0^{-1})\right) \phi^{(\mathrm{II})} \right. \\ & \left. +e^{\beta_+ r_0} \phi^{(\mathrm{I})} \cdot \left(\beta_- + O (r_0^{-1})\right) e^{\beta_- r_0} \phi^{(\mathrm{II})} \right. \\ & \left.+e^{\beta_- r_0} \phi^{(\mathrm{II})} \cdot \left(\beta_+ + O (r_0^{-1})\right) e^{\beta_+ r_0} \phi^{(\mathrm{I})}
  \right]\, .  
  \end{split}
\end{align} Note also, that for $(t,r)$- and $(\tau,r)$-coordinates we have $\beta_+ = - \beta_-$, see e.g. \eqref{eq:betamassive_trFORMAL} and hence the last two lines
in equation \eqref{eq:reg_action_formal}
do not contribute in the limit $r_0 \rightarrow \infty.$

More explicitly, substituting the Fourier space representation \eqref{eq:TRasymptoticFullField} with $\beta_\pm$ given by \eqref{eq:betamassless_tr} for the massless field in $(t,r)$-coordinates, yields
\begin{align}
\begin{split}
\label{eq:regularizedOnShellActionTR}
    S^{\mathrm{reg}}_{\mathrm{os}} =  -\frac{1}{2}\sum_{\ell, I}\,\int_{-\infty}^{\infty} &\, \frac{\dt \omega}{2\pi}  \left[e^{2 i |\omega| r_0}\tilde{\phi}^{(\mathrm{I})}_{0}(-\omega, \ell, I)\tilde{\phi}^{(\mathrm{I})}_{0}(\omega, \ell, I)\left(i |\omega|+O(r_0^{-1})\right)\right. \\  & \left. -e^{-2i |\omega|r_0}\tilde{\phi}^{(\mathrm{II})}_{0}( -\omega, \ell, I)\tilde{\phi}_{0}^{(\mathrm{II})}(\omega, \ell, I)\left(i |\omega|+O(r_0^{-1})\right)   +e^{0} \left(0 + O (r_0^{-1})\right) \right].
     \end{split} 
\end{align}
In the Euclidean version, in $(\tau,r)$-coordinates, using \eqref{eq:EucldieanAsymptoticSeries} and \eqref{eq:betamassless_euc} for $\beta_\pm$, one obtains  the same result with all the factors of $i$ removed. The Euclidean version of \eqref{eq:regularizedOnShellActionTR} is thus exponentially divergent in the $r_0\to \infty$ limit due to the $(\mathrm{I})$ branch (see Appendix \ref{sec:EuclideanOnshellAction} for details), while \eqref{eq:regularizedOnShellActionTR} itself appears to be strongly oscillating as $r_0\to \infty$.

However, the above statement about the Lorentzian integral \eqref{eq:regularizedOnShellActionTR} is more subtle. The crucial underlying reason is that this action should be understood from a Lorentzian path integral perspective, which must be defined  accordingly to the chosen boundary conditions at any finite radius.\footnote{One may be tempted to apply the Riemann-Lebesgue lemma to \eqref{eq:regularizedOnShellActionTR}. However, this would be inconsistent with our problem. The lemma can be applied only if both functions $\phi^{(\mathrm{I})}(t, \Omega)$ and $\phi^{(\mathrm{II})}(t, \Omega)$ are $L^1$-integrable. As discussed further in Section 6, this feature is not generic in scattering and would prevent the definition of functional derivatives of the action with respect to one of the two data.} In order to fix the boundary conditions in a Lorentzian path integral we have to impose an $i\epsilon$-prescription. The boundary conditions in our problem correspond to the scattering data in the asymptotic limit. 

As we are interested in S-matrix elements,
the natural choice for the $i\epsilon$ prescription is 
\begin{align}
\label{eq:iepsilonprescription}
 \omega \mapsto \omega + i \epsilon \ \text{sgn}(\omega) \qquad \textrm{or equivalently} \qquad
    \omega^2 \mapsto \omega^2 + i\epsilon\, ,
\end{align}
used for the Feynman-propagator. Since this $i\epsilon$-prescription implies $|\omega| \mapsto |\omega| + i \epsilon$, we note that $e^{\beta_- r}$ terms diverge for $r \to \infty$ while $e^{\beta_+ r}$ vanishes in that limit. Given the asymptotic expansion of the form \eqref{eq:TRasymptoticFullFieldRepackaged} we refer to $\phi^{\mathrm{(II)}}$ as the leading mode, and to $\phi^{\mathrm{(I)}}$ as subleading mode. 

In stark contrast to AdS, the divergences in the action (the same comment applies to the Euclidean version) are exponential rather than polynomial. This leads to two crucial differences.  First, a naive subtraction of divergences (which is unjustified in any case!)  may result in a vanishing action. Second, there are infinitely many divergent terms that together will form a non-local term. We do not regard this latter fact -- usually seen as problematic -- as overly troublesome. In fact, it might be a key feature of flat holography. However, since this nuance is usually seen with suspicion, let us note a less familiar but important observation concerning the structure of holographic renormalisation in AdS/CFT:  certain interacting scalar fields with irrelevant dimensions do not permit local counterterms at each order in the perturbative expansion \cite{vanRees:2011fr}. It is tantalizing to think that the scalar field in flat spacetime is similar to the irrelevant scalar in AdS/CFT. 

 For the sake of completeness, let us also comment on the Lorentzian on-shell action for scalar fields in $(v,r)$-coordinates. Again, the on-shell action can be written as \eqref{eq:reg_action_formal} with appropriate $\beta_\pm$ given by \eqref{eq:betamassive_vrFORMAL}. Restricting to the massless case, using \eqref{eq:betamassless_vr}, we get
\begin{align}
\begin{split}
\label{eq:regularizedOnShellActionVR}
    S^{\mathrm{reg}}_{\mathrm{os}} = -\sum_{\ell, I}\,\int_{-\infty}^{\infty} \, 
    &\frac{\dt \omega}{2\pi}  \left[e^{2 i |\omega| r_0}\tilde{\phi}^{(\mathrm{I})}( -\omega, \ell, I)\tilde{\phi}^{\mathrm{(I)}}(\omega, \ell, I)\left(i \omega + i |\omega|+O(r_0^{-1})\right)\right. \\ &  \left. +e^{-2i |\omega| r_0}\tilde{\phi}^{(\mathrm{II})}( -\omega, \ell, I)\tilde{\phi}^{(\mathrm{II})}( \omega, \ell, I)\left(i \omega -i |\omega|+O(r_0^{-1})\right)   \right.\\ &\left.+i \omega \left(\tilde{\phi}^{\mathrm{(I)}}( -\omega, \ell, I)\tilde{\phi}^{\mathrm{(II)}}( \omega, \ell, I) + \tilde{\phi}^{\mathrm{(II)}}( -\omega, \ell, I)  \tilde{\phi}^{\mathrm{(I)}}(\omega, \ell, I)  + O (r_0^{-1})\right) \right]. 
     \end{split} 
\end{align} 
The $i \epsilon$ prescription \eqref{eq:iepsilonprescription} modifies \eqref{eq:betamassless_vr} to $\beta_{\pm} = 2(i\omega \mp \epsilon)\theta(\pm \omega)$. As a result, the field exhibits an exponential divergence for negative frequencies and hence also the on-shell action diverges as $r_0\to\infty$. With reference to table \ref{tab:FrequenciesBranches} the oscillating part of the $(\mathrm{II})$-branch becomes exponentially diverging in the $(v,r)$-coordinate system (and similarly in $(u,r)$-coordinates).

In conclusion, all these cases  indicate the need for a procedure that makes the on-shell action and its variation well defined. This is the aim of our holographic renormalization approach, whose general strategy is explained next.

\subsection{Holographic renormalization \& the Hamilton-Jacobi equation}
\label{sec:HJ} 
As mentioned in section \ref{sec:review_holoren}, there are various ways to holographically renormalize the bulk on-shell action in AdS/CFT and get a finite result for the normalized action $S^{\mathrm{ren}}$, a functional of the boundary data. The approach commonly used in AdS, based on an iterative inversion of the asymptotic solutions, fails in flat spacetime where Thomé solutions appear. However, we propose that the Hamilton–Jacobi formalism remains a valid and desirable approach. It avoids the cumbersome inversion of an asymptotic series and ensures by construction that the normalized action is stationary with prescribed boundary conditions.

Let us first discuss the boundary conditions. For the specific scalar system
that we are addressing, the variation can be expressed as 
\begin{equation}
\delta S\,\hat{=}\,\dt \theta(\Phi, \delta \Phi)\,,\qquad [\star\theta(\Phi,\delta\Phi)]^\mu=\sqrt{|g|} \frac{\partial \mathscr{L}}{\partial (\partial_\mu \Phi) }  \delta\Phi\, , 
\end{equation}
where $\hat{=}$ denotes equality on-shell and $\theta$ is the presymplectic potential. The action is thus naturally stationary with Dirichlet conditions $\delta\Phi|_{\Sigma}=0$, $\Sigma$ being the boundary. We take it to be a timelike surface selected by the coordinate $r=r_0$ in an appropriate foliation that approaches $\Sigma_{\infty}=\mathscr{I}^+\cup\mathscr{I^-}$ as $r\to\infty$.\footnote{This is only schematic as we work on the physical spacetime, not the conformally compactified version. Results would correspond. We neglect future or past spacelike surfaces. It would be interesting to explore this further.} We can always find the coordinates \eqref{eq:foliation} adapted to such a foliation,  but the following
considerations can be kept covariant.

As $\Sigma\to\Sigma_{\infty}$ the action is ill defined and explicitly diverges with appropriate $i\epsilon$ prescriptions. The counterterm action to be added to $S$ to form the subtracted action,
\begin{equation}
S^{\mathrm{sub}}=S-S^{\mathrm{ct}}
\end{equation}
is defined on the timelike surface $\Sigma$ by the requirement that the pullback of the subtracted symplectic potential, $\theta^{\mathrm{sub}}$, to $\Sigma$ retains the Dirichlet form and is finite in the asymptotic $\Sigma\to\Sigma_\infty$ limit. Denoting with $\mathcal{F}$ the pullback\footnote{In other words and more in general, the pullback of $\theta$ can be decomposed as $\mathcal{F} - \delta L_b+\dt \mathcal{W}$. The term $\delta L_b$ corresponds to the freedom of selecting boundary conditions and $\dt\mathcal{W}$ is a corner term. For the Klein Gordon action both $L_b$ and $\mathcal{W}$ vanish naturally and $\mathcal{F}$ is automatically of Dirichlet form. The addition of $S^{\mathrm{ct}}$ is required not to change this form. } of $\theta$ to $\Sigma$, e.g. $\mathtt{s}\sqrt{|h|}\partial_n\Phi\delta \Phi$, we thus require 
\begin{equation}
\label{eq:finiteF}
\mathcal{F}^{\mathrm{sub}}=\mathcal{F}-\frac{\delta S^{\mathrm{ct}}}{\delta \Phi}\delta \Phi=:\Pi^{\mathrm{sub}}\delta \Phi \to {O}(1) \quad \text{as} \quad r_0\to\infty\,.
\end{equation}
In the above expression, $\Pi^{\mathrm{sub}}$ is the subtracted radial momentum 
\begin{equation}
\label{eq:def:CanonicalMomentum}
\Pi^{\mathrm{sub}}=\Pi-\frac{\delta S^{\mathrm{ct}}}{\delta \Phi}, \quad \Pi=\frac{\partial L}{\partial \dot{\Phi}}=\mathtt{s}\sqrt{|h|}\partial_{n}\Phi\, . 
\end{equation}
Thus far, we have not specified how to obtain $\Pi^{\mathrm{sub}}$; we will do so shortly, but before that let us make a small jump forward  inspired by the discussion of the previous section. We are explicitly looking for a result of the form
\begin{equation}\label{eq:Fren}
\lim_{r_0\to\infty}\mathcal{F}^{\mathrm{sub}}=\pi\,\delta \phi^{\mathrm{(II)}}\, , 
\end{equation}
where $\pi$ is finite and defined as
\begin{equation}\label{eq:varSren}
\pi=\frac{\delta S^{\mathrm{ren}}}{\delta \phi^{\mathrm{(II)}}}\, , 
\end{equation}
and as we will see explicitly it is related to the mode $\phi^{\mathrm{(I)}}$. Unfortunately, in contrast to AdS/CFT, this is not yet enough to claim that the one-point function of a dual Carrollian operator $\mathcal{O}$ is given by $\pi$. The Carrollian source $\phi_s$  in the asymptotic limit, turns out to be closely related to $\phi^{\mathrm{(II)}}$ consistently with the picture we propose. We revisit this statement in section \ref{sec:holographicCorrlators}.

Now, for the determination of the counterterm, another condition comes handy \cite{deBoer:2000cz,Papadimitriou:2010as}. We require  that the subtracted action is stationary under variations of the regulating surface.\footnote{Together with the previous requirement \eqref{eq:finiteF},  this is the request that the action is stationary under a so-called Weiss variation. This is a neat way to derive the Hamilton-Jacobi formalism bypassing (but equivalent to) the derivation using canonical transformations.} This is equivalent to
\begin{equation}\label{eq:limitdefined}
\lim_{r_0\to\infty}\frac{\dt }{\dt r_0}S^{\mathrm{sub}}=0
\end{equation}
and implies
\begin{equation}
\label{eq:HamiltonJacobiII}
\int \dt^dX\, L[\Phi(r_0), r_0] - \frac{\dt }{\dt r_0} S^{\mathrm{ct}}[\Phi(r_0), r_0]  \overset{r_0 \to \infty}{\longrightarrow} 0\, , 
\end{equation}
where 
\begin{equation}\label{eq:totder}
\frac{\dt }{\dt r} S^{\mathrm{ct}}[\Phi(r), r]   =\int \dt^dX\frac{\delta S^{\mathrm{ct}}}{\delta \Phi}\partial_r \Phi+\partial_r S^{\mathrm{ct}}\,.
\end{equation}
Once we replace
the canonical momentum  in \eqref{eq:HamiltonJacobiII} by $\frac{\delta S^{\mathrm{ct}}}{\delta \Phi}$ using equation \eqref{eq:finiteF}, we obtain the Hamilton-Jacobi equation, which can also be expressed in the form\footnote{\label{footnote:Replacement}One might be concerned about finite or divergent terms when this replacement is performed. However, in the theories we consider, including those with interaction terms of the form $\Phi^n$, this replacement does not change the asymptotic behavior of \eqref{eq:HamiltonJacobiII}. }
\begin{equation}
\partial_{r_0}S^{\mathrm{ct}}+H\left(\frac{\delta S^{\mathrm{ct}}}{\delta \Phi}, \Phi(r_0), r_0\right)\to 0\, . 
\end{equation}
The Lagrangian density $\mathscr{L}$ for the case of our example in the ADM coordinates reads
\begin{equation}
\mathscr{L}= \frac{\mathtt{s}}{2}\, N\sqrt{|h|}
\left(\frac{\mathcal{V}^{2}}{N^{2}}+ h^{ab}\partial_{a}\Phi\,\partial_{b}\Phi+ m^{2}\Phi^{2}\right),\qquad \mathcal{V}:= \dot{\Phi}-N^{a}\partial_{a}\Phi= N\,\partial_{n}\Phi\,,
\end{equation}
and introducing $\tilde{\Pi} := \Pi/(\mathtt{s}\sqrt{|h|})$, the Hamiltonian density is $H = \mathtt{s}\sqrt{|h|}(N\mathscr{H} + N^{a}\mathscr{H}_{a})$ where 
\begin{equation}
\mathscr{H}= \frac{1}{2}\left(\tilde{\Pi}^{2}-h^{ab}\partial_{a}\Phi\,\partial_{b}\Phi-m^{2}\Phi^{2}\right),\qquad \mathscr{H}_{a}= \tilde{\Pi}\,\partial_{a}\Phi .
\end{equation}
This splitting is fictitious in a fixed background, but it would be promoted to the Hamiltonian and momentum constraints when gravity is dynamical. The Hamilton-Jacobi equation in its ``Lagrangian form" \eqref{eq:HamiltonJacobiII} takes the form 
{\small
\begin{align}\label{eq:HamiltonJacobi_general}
\begin{split}
\int \dt^d X&\left\{\sqrt{|h|}N\frac{\mathtt{s}}{2}\left[\left(\frac{1}{\sqrt{|h|}}\frac{\delta S^\mathrm{ct}}{\delta \Phi}\right)^2-(h^{ab}\partial_a\Phi\partial_b\Phi+m^2\Phi^2)\right]+\frac{\delta S^\mathrm{ct}}{\delta h_{ab}}\partial_r h_{ab}+\frac{\delta S^\mathrm{ct}}{\delta \Phi}N^a\partial_a\Phi\right\}=0\,.
\end{split}
\end{align}}
where we expressed \eqref{eq:totder} via
\begin{equation}
\frac{\dt }{\dt r} S^{\mathrm{ct}}=\int \dt^dX\left(\frac{\delta S^{\mathrm{ct}}}{\delta \Phi}\partial_r \Phi+  \frac{\delta S^{\mathrm{ct}}}{\delta h_{ab}}\partial_r h_{ab}\right)\,,
\end{equation}
noticing that the only explicit dependence of $S^{\mathrm{ct}}$ on $r$ is through the induced metric.\footnote{In the AdS/CFT correspondence, in the presence of anomalies there are explicitly $r$ dependent terms in $S^{\mathrm{ct}}$. These are unavoidable if locality of the counterterm action is preserved. In our flat case, locality does not hold anymore. It remains an open question to assess the existence of quantum anomalies in flat holography.} While this equation is derived for the massive free field, an immediate generalisation is the case of a self-interacting scalar with potential interactions. This simply amounts to the replacement of $m^2\Phi^2$ with $U(\Phi)=m^2\Phi+2V(\Phi)$.

\section{Holographic renormalization of the free scalar action}\label{sec:Holographicrenormalizationformasslessfields}

This section presents the details of the construction of the counterterm according to the previous strategy. We restrict to $N=1$ in accordance with the two gauges we have selected for the analysis of this paper. To solve the free Hamilton-Jacobi equation, we use a counterterm ansatz of the form 
\begin{align}\label{eq:AnsatzCounterterms}
    S^{\mathrm{ct}} = -\frac{1}{2} \int_{r= r_0} \dt^d X\, \sqrt{|h|}\, \Phi \,f\, \Phi\,,
\end{align}
where $f$ is a symmetric differential operator on the slice. Inserting  \eqref{eq:AnsatzCounterterms} into \eqref{eq:HamiltonJacobi_general} we get\footnote{We use the symmetry of $f$ as  $\int \dt^d X \sqrt{|h|} \,\Phi_1 \,(f \Phi_2) =\int \dt^d X \sqrt{|h|}\, (f \Phi_1)\, \Phi_2 $ for all $\Phi_{1,2}$. } 
\begin{equation}
\label{eq:GeneralRiccati}
\int_{r = r_0} \dt^d X\left\{\sqrt{|h|}\mathtt{s}\left[\frac{1}{2}\Phi\left(f^2+\Box_h-m^2\right)\Phi\right]+\frac{\delta S^{\mathrm{ct}}}{\delta h_{ab}}\partial_r h_{ab}+\sqrt{|h|}f\Phi N^a\partial_a\Phi\right\}=0\, . 
\end{equation}
The ansatz \eqref{eq:AnsatzCounterterms}
is reminiscent of the AdS counterterm, but as we detail in the next subsections, we are forced to consider a non-covariant split on $r=const$ because, for example, the time and spatial part of $\Box_h$ scale differently as $r\to\infty$. We will accept this and refer to the discussion. 

Section \ref{sec:holoren_tr} and \ref{sec:holoren_vr} deal with the massless scalar in $(t,r)$- and $(v,r)$-coordinates, respectively.  Section \ref{sec:holographicCorrlators} presents the holographic two-point function. Section \ref{sec:Massive} comments on the case of a massive scalar. 

\subsection{Renormalization in $(t,r)$-coordinates}\label{sec:holoren_tr}
The metric in $(t,r)$-coordinates is the special case of \eqref{eq:foliation} with
\begin{equation}
N=1,\quad  N^a =0,\quad   h_{tt}=-1,\quad  h_{ti}=0,\quad  h_{ij}=\gamma_{ij}=r^2\hat{g}_{ij}
\end{equation}
where $\hat{g}$ is a generic metric on the $(d-1)$-dimensional sphere, such that $R_{ij}[\gamma] = \frac{d-2}{r^2}\gamma_{ij}$. For notational convenience,
\begin{align}
    \dt s^2 &=  \dt r^2 - \dt t^2 + \gamma_{ij}(r,\vartheta)\dt \vartheta^i \dt \vartheta^{j} = \dt r^2 - \dt t^2 + r^2\hat{g}_{ij}(\vartheta)\dt \vartheta^i \dt \vartheta^{j}\, , 
\end{align}
and the free, massless scalar field Lagrangian reads as
\begin{align}
   \mathscr{L} = -\frac{1}{2} \left(\dot{\Phi}^2 - \left(\partial_t\Phi\right)^2 + \gamma^{ij}\partial_i\Phi\partial_j \Phi\right)\,, 
\end{align}

In order to express equation \eqref{eq:GeneralRiccati} in a partially covariant way, as discussed at the beginning of this section, 
we take the symmetric function $f$, which characterises the ansatz \eqref{eq:AnsatzCounterterms}, to be a  function of the even time derivative $-\partial_t^2$, the Laplace operator on the sphere $\Delta_{\gamma}$, and the Ricci scalar of $\gamma$, $R[\gamma]$, using the identity \eqref{eq:identities1} to express $r$ in terms of $R[\gamma]$. In this way we hide the radial dependence in covariant objects (see also \cite{Papadimitriou:2010as}). Using also the identities \eqref{eq.var_identities} to express the variations of $S^{\mathrm{ct}}$,  the Hamilton-Jacobi equation becomes 
\begin{align}
    \begin{split}
    \frac{1}{2}\int_{r=r_0} \dt^d X\, \sqrt{\gamma}\, \Phi \left[\phantom{\frac{\phantom{j}}{}}\right. &\hspace{-0.2cm}f^2 -( \partial_t^2 - \Delta_{\gamma}) \\
    \,+ &\left.
   \sqrt{\frac{4 R[\gamma]}{(d-1)(d-2)} }\left(\frac{d-1}{2} f - R[\gamma] 
 \frac{\partial f}{\partial R[\gamma]}  - \Delta_{\gamma} \frac{\partial f}{\partial \Delta_{\gamma}}\right)\right]\Phi =0 \, . 
    \end{split}
\end{align}
Integrating by parts (recall that the derivative operators in $f$ are symmetric) and neglecting the boundary terms according to our previous assumptions, the left hand side becomes
\small
\begin{align}
\label{eq:HJTRintermediate}
 \frac{1}{2}\int_{r=r_0} \dt^d X\, \sqrt{\gamma}\, \Phi \left[f^2 -( \partial_t^2 - \Delta_{\gamma})+ 
   \sqrt{\frac{R[\gamma]}{(d-1)(d-2)} }\left((d-1) f - 2R[\gamma] 
 \frac{\partial f}{\partial R[\gamma]}  - 2\Delta_{\gamma} \frac{\partial f}{\partial \Delta_{\gamma}}\right)\right]\Phi  \,  
\end{align}
\normalsize
which we set to vanish in the large $r$ limit. 

Schematically we can make the following argument.\footnote{Recall that in time domain $\beta_\pm$ are differential operators, so the following statement is made precise by simply rephrasing it in frequency space.}  The leading mode of $\Phi$ is $\Phi \sim e^{\beta_-r}$ (recall the analytic continuation), hence the term in the parenthesis of \eqref{eq:HJTRintermediate} should falloff faster than $e^{-2\beta_-r}$.  A way to ensure this is simply to impose that $f$ is an asymptotic solution of the differential equation
\begin{equation}\label{eq:hamilton_jacobi_eq_tr}
f^2 -( \partial_t^2 - \Delta_{\gamma})+ 
   \sqrt{\frac{R[\gamma]}{(d-1)(d-2)} }\left((d-1) f - 2R[\gamma] 
 \frac{\partial f}{\partial R[\gamma]}  - 2\Delta_{\gamma} \frac{\partial f}{\partial \Delta_{\gamma}}\right)=0\,.
\end{equation}
We can write this equation as a standard ordinary differential equation, via change of variables  similar to those  performed in \cite{Papadimitriou:2010as}. A faster way to obtain the same result is by changing back to $r$. So we can write it more compactly, at the price of not having it manifestly covariant, as 
\begin{align}
\label{eq:RiccatiEquation}
\partial_r f+\frac{d-1}{r}f + f^2  -  \left(\partial_t^2 - \frac{1}{r^2}\Delta_{\hat{g}}\right)=0\, . 
\end{align}
The Hamilton-Jacobi equation thus reduces to a non-linear  Riccati differential equation for $f$ and is related to the factorization of the equations of motion as reviewed in Appendix \ref{sec:factorization} (see also \cite{Papadimitriou:2003is,Papadimitriou:2013jca}).

The solution of \eqref{eq:RiccatiEquation} can be obtained with standard techniques by transforming it to an ordinary linear differential equation via the substitution
\begin{equation}\label{eq:definition_f}
f= \partial_r \log g\,.
\end{equation}
The resulting equation for $g = g(r, -\partial_t^2, \Delta_{\hat{g}})$ is the Klein-Gordon equation in the given coordinate system
\begin{align}
    \left(\partial_r^2 +\frac{d-1}{r}\partial_r +\frac{1}{r^2}\Delta_{\hat{g}} -\partial_t^2\right)g =0\, .
\end{align}
Hence $g$ takes the form 
\begin{align}
\label{eq:DefginCounterterms}
    g = \frac{1}{r^{\frac{d-1}{2}}}\left[ e^{\beta_+ r}\sum_{k=0}^{\infty} r^{- k}C^{(\mathrm{I})}_k + e^{\beta_- r}\sum_{k=0}^{\infty} r^{- k}C^{(\mathrm{II})}_k\right]\, , 
\end{align} 
where $\beta_{\pm}$ were defined as the massless limit of \eqref{eq:betamassive_trFORMAL} and $C^{(\mathrm{I}), \, (\mathrm{II})}_k$ are functions of $(-\partial_t^2, \Delta_{\hat{g}})$, respectively determined for $k\ge 1$ by the functions $C^{\mathrm{(I)}, \, \mathrm{(II)}}_0$. On the other hand, the function 
$f$ should only depend on one integration function. We can express it as the ratio of the two functions, $C^{(\mathrm{I})}_0(-\partial_t^2, \Delta_{\hat{g}})$ and $C^{(\mathrm{II})}_0(-\partial_t^2, \Delta_{\hat{g}})$. A convenient form is 
\begin{align}\label{eq:ratio}
    {C}(-\partial_t^2, \Delta_{\hat{g}}) = \frac{C^{(\mathrm{I})}_0}{C^{(\mathrm{II})}_0}\, . 
\end{align}
In \eqref{eq:ratio} we clearly assume that  $C^{(\mathrm{II})}_0$ is non vanishing (actually, the operator $C^{(\mathrm{II})}_0$  is invertible). This is indeed an appropriate choice because if $C^{\mathrm{(II)}}_0$ vanishes, the counterterm action \eqref{eq:AnsatzCounterterms} would behave asymptotically as
\begin{align}
S^{\mathrm{ct}}= \int_{r=r_0} \dt^d X \sqrt{\gamma}\,\Phi(\partial_r\log g)\Phi   \sim \beta_+ e^{2 \beta_- r_0} \phi^{\mathrm{(II)}}\phi^{\mathrm{(II)}}   + \dots\, , 
\end{align}
which does not match the divergences of the regulated on-shell action \eqref{eq:reg_action_formal}, which we report here schematically for convenience
\begin{align}\label{eq:divergence}
S_{\mathrm{os}}^{\mathrm{reg}}   \sim \beta_-e^{2 \beta_- r_0} \phi^{\mathrm{(II)}}\phi^{\mathrm{(II)}}   + \dots\, . 
\end{align}
We remark that this argument only shows the necessity of $C_{0}^{\mathrm{(II)}}\neq 0$, so that the function $f$ is of the form\footnote{\label{Hankelfootnote}We do not require the solution to be well defined in the deep interior of the bulk, which would relate the two asymptotic branches. For pure Minkowski spacetime, we can write the function $f$ in terms of the expansion that is known from section \ref{sec.FrobeniusThome}. In this case we can also express it in terms of Hankel functions 
$ f = \partial_r \log(g)= \partial_r \log\left(r^{-\frac{d-1}{2}}\,\left(H^{(2)}_{\nu}(|\omega|r) + C H^{(1)}_{\nu}(|\omega|r) \right)\right)$. This implies that the ellipsis in \eqref{eq:finTR} also contain $C$-dependent subleading terms. 
} 
\begin{align}
\label{eq:finTR}
    f = \partial_r \log(g)= \partial_r \log(r^{-\frac{d-1}{2}}e^{\beta_- r} + \dots) = \beta_- -\frac{d-1}{2r} +\dots\,.
\end{align}
and hence
\begin{align}
S^{\mathrm{ct}}= \int_{r=r_0} \dt^d X \sqrt{\gamma}\,\Phi(\partial_r\log g)\Phi   \sim \beta_- e^{2 \beta_- r_0} \phi^{\mathrm{(II)}}\phi^{\mathrm{(II)}}   + \dots\, .
\end{align}
Note that this argument works for any $C(\partial_t, \Delta_{\hat{g}})$, provided that $C_0^{\mathrm{(II)}}$ is non-zero. From now on, we consider the case $C = 0$ and we comment on its generalizations later.

With the given counterterm, the subtracted action and momentum are
\begin{equation}
  S^{\mathrm{sub}} = -\frac{1}{2} \int_{r=r_0} \dt^d X \sqrt{\gamma} \,\Phi \left(\partial_r -f\right)\Phi, \qquad \Pi^{\mathrm{sub}}=-\sqrt{\gamma}\left(\partial_r -f\right)\Phi
\end{equation}
It is important to point out that we have not yet explicitly demonstrated that they stay finite in the large-$r$ limit. As discussed in section \ref{sec:need_holo}, there is in fact an infinite number of divergences, so the proof cannot proceed by simple iteration. Nevertheless, the essential ingredients of the procedure already contain the proof. Let us now make this explicit by writing the asymptotic solution $\Phi$ -- determined by the data $\phi^{\mathrm{(I),(II)}}$ -- in terms of $g$, which is determined by $C_0^{\mathrm{(I),(II)}}$. To simplify the expression, having already selected $C=0$, we can fix without loss of generality $C_0^{\mathrm{(II)}} = 1$. Thus, 
\begin{align}
    \Phi  = g \,\phi^{\mathrm{(II)}}  + \left(r^{-\frac{d-1}{2}}e^{\beta_+r} + \dots\right)\phi^{\mathrm{(I)}}. 
\end{align}
Since by construction of $f$, $(\partial_r-f)g=0$, we are left with 
\begin{align}
  (\partial_r-f)\Phi = (\beta_+ - \beta_-)\left(r^{-\frac{d-1}{2}}e^{\beta_
    +r} + \dots\right)\phi^{\mathrm{(I)}}. 
\end{align}
We thus obtain that 
\begin{equation}
  \Pi^{\mathrm{sub}} = -\sqrt{\gamma}(\beta_+ - \beta_-)r_0^{-\frac{d-1}{2}}\left(e^{\beta_
    +r_0} + \dots\right)\phi^{\mathrm{(I)}} 
\end{equation}
which vanishes as $r_0\to\infty$ as required for $\Pi^{\mathrm{sub}}$. We thus conclude that 
\begin{equation}
  S_{\mathrm{os}}^{\mathrm{ren}} = \lim_{r_0 \to \infty}  S_{\mathrm{os}}^{\mathrm{sub}}= \lim_{r_0 \to \infty}  \frac{1}{2}\int_{r=r_0} \dt^d X \, \Phi \Pi^{\mathrm{sub}} 
\end{equation}
is formally 
\begin{equation}
 S_{\mathrm{os}}^{\mathrm{ren}} = -\frac{1}{2}\int \dt^d X \sqrt{\hat{g}} \,\phi^{(\mathrm{II})} (\beta_+ - \beta_-)\, \phi^{(\mathrm{I})} \,, 
\end{equation}
which is finite, as we wanted to show. 
We can evaluate it explicitly in frequency space
\begin{align}
\label{eq:SrenTRfrequencyspace}
    S_{\mathrm{os}}^{\mathrm{ren}} &= -\frac{i}{2\pi} \int \dt \Omega \,  \int_{\mathcal{C}} \dt \omega\, \tilde{\phi}^{(\mathrm{II})}(-\omega) |\omega| \tilde{\phi}^{(\mathrm{I})}(\omega)\,. 
\end{align}
The $\mathcal{C}$ indicates that, due to the $i\epsilon$ prescription, we are not integrating over the real line but over a deformed contour. 
Transforming back to position space, we get 
\begin{align}
    S_{\mathrm{os}}^{\mathrm{ren}} &=-\frac{i}{2 \pi} \int \dt \Omega\,  \int \dt t_1 \, \dt t_2 \, \phi^{(\mathrm{II})}(t_1, \Omega) \phi^{(\mathrm{I})}(t_2, \Omega) \, \int_{\mathcal{C}} \dt \omega e^{-i \omega (t_1 - t_2) }|\omega| \, .  
\end{align}
We can make the contour manifest again by inserting the $i\epsilon$ prescription \eqref{eq:iepsilonprescription}, or,
    $\omega \mapsto \omega + i \omega \epsilon$,
leading effectively to the same contour prescription. The frequency integral yields
\begin{align}
\label{eq:SrenTRpositionspace}
    S_{\mathrm{os}}^{\mathrm{ren}} =  \frac{i}{\pi} \int \dt \Omega \,  \int \dt t_1 \, \dt t_2 \, \frac{1}{(t_1 - t_2)^2-i\epsilon}\phi^{(\mathrm{II})}(t_1, \Omega)\phi^{(\mathrm{I})}(t_2, \Omega)\, . 
\end{align} 
Terms with $\phi^{(\mathrm{I})}\phi^{(\mathrm{I})}$ are suppressed due to the $i\epsilon$ prescription as they are multiplied by $e^{\beta_+ r}$ and are therefore not present in the limit. 

The relation of the modes in $(t, r)$-coordinates to modes on $\scri^+$ and $\scri^-$, and hence to scattering data, is not obvious. Therefore, we repeat the analysis in ingoing $(v.r)$-coordinates.

\subsection{Renormalization in $(v,r)$-coordinates}
\label{sec:holoren_vr}

The spacetime metric in $(v,r)$-coordinates is obtained from the ADM form via the choice 
\eqref{eq:HamiltonJacobi_general} with \begin{align}
N^2=1, \quad N^{v}=-1, \quad N^i=0, \quad
h_{v v}=-1, \quad h_{v i} =0 , \quad h_{ij}=\gamma_{ij}\, . 
\end{align}
The canonical momentum \eqref{eq:def:CanonicalMomentum} reads in this coordinate system\footnote{Note that the term depending on $\partial_v \Phi$ contributes as a codimension-two term in the on-shell action but not in $\delta S$.}
\begin{align}
    \Pi = -\sqrt{\gamma}\left(\partial_r + \partial_v\right)\Phi\, . 
\end{align}
The function $f$ in \eqref{eq:AnsatzCounterterms}, 
will be now taken to depend on $r, \, -\Delta_{\mathds{S}^{d-1}}$ and $-\partial_v^2$. Following a  procedure similar to the one detailed before, the Hamilton-Jacobi equation takes the form
\begin{align}
\label{eq:RiccatiVR}
    -\frac{1}{2}\int_{r=r_0} \dt^d X \, \sqrt{\gamma}\, \Phi\left(f^2 - \partial_v^2 +\frac{1}{r^2}\Delta_{\mathds{S}^{d-1}} + \frac{d-1}{r}f + \partial_r f \right)\Phi =0\, , 
\end{align}
where we neglected corner terms. 
This leads to a Riccati equation which resembles  \eqref{eq:RiccatiEquation} with $\partial_t$ exchanged by $\partial_v$. Therefore, the solution of the Riccati equation \eqref{eq:RiccatiVR} agrees with \eqref{eq:finTR}, where $\beta_{\pm}$ are still taken from \eqref{eq:betamassive_trFORMAL} with $\partial_t\to\partial_v$. 
The combination $\partial_r+ \partial_v - f$ will indeed cancel the diverging branch that behaves as $e^{(i\omega - i|\omega|)r}r^{-\frac{d-1}{2}}+\dots\,$. 
Thus, we are left with the canonical momentum \eqref{eq:varSren} of the form
\begin{align}
    \pi=-\sqrt{\hat{g}}\left(\beta_+ - \beta_-\right)\phi^{\mathrm{(I)}}\, , 
\end{align}
with $\beta_{\pm}$ defined in \eqref{eq:betamassless_vr}. The renormalized on-shell action in frequency space reads
\begin{align}
\label{eq:SOnShellTRfreq}
\begin{split}
    S^{\mathrm{ren}}_{\mathrm{os}} =& -\frac{1}{4 \pi}\int \dt \omega\, \dt \Omega\, \tilde{\phi}^{(\mathrm{II})}(- \omega) (\beta_+ - \beta_-)\tilde{\phi}^{(\mathrm{I})}(\omega)\\=&- \frac{i}{2 \pi }\int \dt \omega\, \dt \Omega\, \tilde{\phi}^{(\mathrm{II})}(-\omega)\,|\omega |\, \tilde{\phi}^{(\mathrm{I})}(\omega)\, . 
    \end{split}
\end{align}
As expected, this agrees with the computation in $(t,r)$-coordinates \eqref{eq:SrenTRfrequencyspace}. In position space we would naturally write it as
\begin{align}
\label{eq:SrenVRpositionspace}
    S_{\mathrm{os}}^{\mathrm{ren}} = \frac{i}{ \pi} \int \dt \Omega \,  \int \dt v_1 \, \dt v_2 \, \frac{1}{(v_1 - v_2)^2 - i \epsilon}\phi^{(\mathrm{II})}(v_1, \Omega)\phi^{(\mathrm{I})}(v_2, \Omega)\,.
\end{align}
The renormalized on-shell action $S^{\mathrm{ren}}_{\mathrm{os}}$ is a functional that depends on the function $\phi^{\mathrm{(II)}}$. 
Also in this computation we have used the choice $C(\partial_v, \Delta_{\hat{g}}) = 0$.

According to \eqref{eq:IntroDictionary}, we identify $S^{\mathrm{ren}}_{\mathrm{os}}$ with the generating functional of connected diagrams, which can be used to compute Carrollian correlation functions as explicitly shown in the next section.

\subsection{Holographic correlators}
\label{sec:holographicCorrlators}

By construction, or by explicit application of the fundamental identity of variational calculus and  knowing that $f$ is symmetric,
\begin{equation}
\delta S^{\mathrm{sub}} = 0 \quad \text{on-shell, with} \quad \delta \Phi = 0 \, ,
\end{equation}
which translates in the asymptotic phase space to the fixing  of $\phi^{\mathrm{(II)}}$. Indeed 
\small
\begin{align}
    \delta S^{\mathrm{sub}} =& - \int_{r=r_0} \dt^d X\, \sqrt{\gamma}\,r^{-\frac{d-1}{2}} \left(\partial_r - f \right) \Phi \left(\left(e^{\beta_- r} + \dots\right)\delta \phi^{\mathrm{(II)}}+\left(e^{\beta_+ r} + \dots\right)\delta \phi^{\mathrm{(I)}} \right) \nonumber\\
    =&- \int_{r=r_0} \dt^d X\, \sqrt{\hat{g}}\, \left((\beta_+ - \beta_-) e^{\beta_+ r} \phi^{\mathrm{(I)}}+\dots\right) \left(\left(e^{\beta_- r} + \dots\right)\delta \phi^{\mathrm{(II)}}+\left(e^{\beta_+ r} + \dots\right)\delta \phi^{\mathrm{(I)}} \right)\\
    =&- \int_{r=r_0} \dt^d X \sqrt{\hat{g}}\,(\beta_+ - \beta_-)\, \phi^{\mathrm{(I)}}\delta \phi^{\mathrm{(II)}} + \dots\, ,  \nonumber
\end{align}
\normalsize
where the ellipsis do not survive the $r_0 \to \infty$ limit. 
This shows that in the large $r_0$ limit, the variation of the renormalized on-shell action vanishes once $\delta\phi^{\mathrm{(II)}}=0$. 
We conclude that the bulk quantity that we associate with the source on the Carrollian side is proportional to $\phi^{\mathrm{(II)}}$, so that the scattering boundary conditions are correctly implemented as evident from the discussion around table \ref{tab:FrequenciesBranches}.

Let us for the moment identify $\phi^{\mathrm{(II)}}=:\phi_s$, where $\phi_s$  is the source of the dual Carrollian operator $\mathcal{O}$ so that the generating functional defined by
\begin{equation}\label{eq:CarrW}
W_{\mathrm{CarrCFT}}[\phi_s]= \log \, \left\langle\exp\left(\int \dt v \,\dt \Omega\, \phi_s(v,\Omega)\mathcal{O}(v,\Omega)\right)\right\rangle\,,
\end{equation}
and assume that this is equal to the renormalized bulk on-shell action. Hence, the Carrollian one-point function is obtained by taking one functional derivative of the renormalized on-shell action with respect to  $\phi^{\mathrm{(II)}}$
\begin{align}
\label{eq:generalOnePointFunction}
    \langle \mathcal{O}_{\Phi}\rangle_{\phi^{\mathrm{(II)}}} = \lim_{r_0 \to \infty}\frac{1}{\sqrt{\hat{g}}}\frac{\delta}{\delta \phi^{\mathrm{(II)}}} S^{\mathrm{sub}}_{\mathrm{os}}[r_0, \phi^{\mathrm{(II)}}]\, . 
\end{align}
Explicitly we have,
\begin{align}
    \langle  \mathcal{O}_{\Phi} (v, \Omega)\rangle_{\phi^{\mathrm{(II)}}} &= \frac{i}{\pi}\, \int \dt v' \frac{1}{(v-v')^2 - i \epsilon} \phi^{\mathrm{(I)}}(v', \Omega)\, \label{eq:OnepointFunctionV}\\
    \langle  \mathcal{O}_{\Phi} (\omega, \Omega)\rangle_{\phi^{\mathrm{(II)}}} &= -\frac{i}{2 \pi}\, |\omega|\tilde{\phi}^{\mathrm{(I)}}(-\omega, \Omega)\,. \label{eq:onepointfunction}
\end{align}

Recall that for this result we have set the function $C$ in \eqref{eq:ratio} to zero. A non-vanishing $C$ adds a finite term of the form $\sim C\,\phi^{\mathrm{(II)}}$ to the one-point function. 
The choice to set $C = 0$ is unique in the sense that $\phi^{\mathrm{(I)}}$ relates to the one-point function while $\phi^{\mathrm{(II)}}$ relates to the source. For a discussion of this choice from the perspective of symplectomorphisms, see \cite{Papadimitriou:2010as}. 

The computation of the one-point function \eqref{eq:generalOnePointFunction} does not require regularity in the interior of the spacetime. However, in order to compute higher point correlation functions, we have to impose regularity. 
In the case of pure Minkowski spacetime, the regularity of the solutions, discussed at the level of \eqref{eq:regularityInterior} and \eqref{eq:TranslatingAsympVSExact}, implies
\begin{align}
\label{eq:regularityForSources}
    \phi^{(\mathrm{I})} = e^{ -i \left( \nu \pi + \frac{\pi}{2}\right)}\phi^{(\mathrm{II})}\, , 
\end{align} 
where $\nu$ was defined in 
 \eqref{eq:nuDef} as an operator acting on the spherical harmonics, with eigenvalues \eqref{eq:EigenvalueNU}. 

Applying a further functional derivative of the one-point function \eqref{eq:OnepointFunctionV}, and using the regularity condition, the Carrollian two point function would take the form

\begin{align}
\begin{split}
\label{eq:phi2Carollian2ptfunctionTR}
    \left.\frac{1}{\sqrt{\hat{g}}}\frac{\delta}{\delta \phi^{\mathrm{(II)}}(v_1)}\langle \mathcal{O}(v_2)\rangle_{\phi^{\mathrm{(II)}}}\right|_{\phi^{\mathrm{(II)}} =0}  =  \frac{1}{ \pi} \frac{e^{-\frac{i\pi (d-2)}{2}}}{(v_1 - v_2)^2 -i \epsilon} \delta^{d-1}(\Omega_1 - \Omega^{\mathrm{AP}}_2)\, ,  
\end{split}
\end{align}
where $\Omega_2^{\mathrm{AP}}$ is antipodally related to $\Omega_2$. The delta function is to be understood as \eqref{eq:AngularDelta}. 
This two-point function matches the two-point functions obtained in  \cite{Kraus:2024gso,Kraus:2025wgi}. However, in order to present it in a form that corresponds exactly to the two-point function of an electric Conformal Carrollian theory, e.g.  $\sim \delta (\Omega_1 - \Omega_2)$, we need to redefine our original definition of sources. This is indeed a natural thing to do because in field theories we expect the correlators to depend on the difference of spacetime points. 
Therefore, we revisit our tentative definition of the source.

As explained in section \ref{sec:SingularitiesOfDifferentialEquations}, $\phi^{\mathrm{(II)}}$ contains both, the positive frequency modes on $\scri^{-}$ and the negative frequency modes on $\scri^{+}$. 
Note however, that we did not take into account that we have to antipodally identify the spheres on $\scri^{-}$ and $\scri^{+}$. Therefore, the appropriate choice of source is 
\begin{align}
\label{eq:defPhiSviaoperator}
\begin{split}
    \tilde{\phi}_s(\omega, \hat{x}) =&  \tilde{\phi}^{\mathrm{(II)}}(\omega, \mathrm{sign}(\omega)\hat{x}) =\begin{cases}
        \tilde{\phi}^{\mathrm{(II)}} (\omega, \hat{x}) & \omega >0 \\
        \tilde{\phi}^{\mathrm{(II)}} (\omega, -\hat{x}) & \omega <0 \\
    \end{cases}
    \end{split}\, , 
\end{align}
where we expressed the angular part as the unit vector $\hat{x}$. It is clear that the counterterms and hence the renormalized on-shell action are not affected by this redefinition. Rewritten in terms of spherical harmonics with the notation introduced in \eqref{eq:TRasymptoticFullField}, the source takes the form
\begin{align}
    \tilde{\phi}_s(\omega, \ell, I) &= \left(\frac{\omega}{|\omega|}\right)^{\ell} \tilde{\phi}^{(\mathrm{II})}(\omega,\ell,I) \, . 
\end{align}

Following the same logic of \eqref{eq:generalOnePointFunction} with, now, $\phi_s$ instead of $\phi^{\mathrm{(II)}}$, we obtain for the free scalar field in asymptotically Minkowski space time the one-point function 
\begin{align}
\label{eq:one-point-function}
    \langle  \mathcal{O}_{\Phi} (\omega, \hat{x})\rangle_{\phi_s} &= -\frac{i}{2 \pi}\, |\omega|\tilde{\phi}^{\mathrm{(I)}}(-\omega, \mathrm{sign}(\omega)\hat{x}) \, . 
\end{align}
Restricting ourselves to pure Minkowski spacetime and demanding regularity of the field solution everywhere by imposing \eqref{eq:regularityForSources}, we can compute the second derivative of the on-shell action
\begin{align}
\begin{split}
\label{eq:Carollian2ptfunctionTR}
\langle \mathcal{O}(v_1) \mathcal{O}(v_2) \rangle =
    \frac{1}{\sqrt{\hat{g}}}\frac{\delta}{\delta \phi_s(v_1)}\left.\langle \mathcal{O}(v_2)\rangle_{\phi^{\mathrm{(II)}}}\right|_{\phi_s =0}   =  \frac{1}{ \pi} \frac{e^{-\frac{i\pi (d-2)}{2}}}{(v_1 - v_2)^2-i\epsilon} \delta^{d-1}(\Omega_1 - \Omega_2)\, ,  
\end{split}
\end{align}
This is our final result for the conformal Carrollian two-point function. If we identify the scaling dimension of our Carrollian operators as 
\begin{equation}
\label{eq:scalaingDimension}
\Delta=\frac{d+1}{2},
\end{equation}
our two-point function agrees with the existing results \cite{Kulkarni:2025qcx, Nguyen:2023miw, Donnay:2022wvx, Mason:2023mti, Kraus:2024gso}. 
Notice that \eqref{eq:scalaingDimension} may seem odd. In fact, the scaling dimension of the Carrollian operator in the literature is $\Delta_\mathrm{there} =\frac{d-1}{2}$, corresponding to the scaling dimension of a massless scalar in $(d+1)$ spacetime dimensions. 
This apparent discrepancy is due to the different definition of Carrollian sources in those papers compared to our prescription in \eqref{eq:onepointfunction}. In particular, the factor of  $|\omega|$ in \eqref{eq:onepointfunction} is ultimately responsible for the discrepancy; the factor of $|\omega|$ arises since we consider $\phi_s$ (related to $\phi^{\textrm{(II)}}$) as the source rather than $\phi_2$ defined by \eqref{eq:Formal_massive_tr_solution}.\footnote{Note that in conformal Carrollian field theories the $v$-derivative of a primary operator is again primary.} Our definition of the source, in terms of $\phi^{\mathrm{(II)}}$, is the exact parallel of the way sources are defined in AdS/CFT: explicitly, in AdS/CFT it is $\phi_{(\Delta_-)}$ in \eqref{eq.solutionAdS} that is identified with the source of the dual operator, not $\phi_1$ given by \eqref{eq:exactAdSsolution}. 

Few remarks are in order now. Although the computation is presented in the $(v,r)$-coordinate system, this result should not be interpreted as the projection of a bulk two-point propagator to $\mathscr{I}^-$. In frequency space it is tempting to consider one of the dual Carrollian operators being inserted at $\scri^-$ and the other at $\scri^+$, since the source $\phi_s$ encodes the scattering data. However, \eqref{eq:Carollian2ptfunctionTR} does not manifestly encode this information in position space. 
We rather advocate for a different interpretation: the Carrollian manifold, on which the putative dual theory is defined, is not identified with either $\scri^{+}$ or $\scri^{-}$. On the right hand side of \eqref{eq:Carollian2ptfunctionTR},  $(v_1,\Omega_1)$ and $(v_2, \Omega_2)$ denote the points on an abstract Carrollian manifold. The coordinate $v$ is now interpreted as the degenerate direction of the abstract Carrollian manifold.

This discussion supports what we advertised in the introduction. The conformal Carrollian theory -- conjecturally dual to an asymptotically flat quantum-gravity theory, of which this fixed background scalar might constitute a sector -- is formulated on a Carrollian manifold  that, a priori, need not be identified with $\scri^{+}$ or $\scri^{-}$. Its only connection to null infinity lies in the shared isometries.

\subsection{Massive Fields}
\label{sec:Massive}
Our approach to compute a renormalized action, can also be applied to massive fields. We briefly show this below, although the Carrollian interpretation of the results is still obscure. 

Let us consider the problem in $(v,r)$-coordinates, so to parallel section \ref{sec:holoren_vr}. Using the asymptotic expansion determined in section \ref{sec.FrobeniusThome}, with \eqref{eq:BetaMassiveVR} supplemented with the $i\epsilon$ prescription \eqref{eq:iepsilonprescription}, namely 
\begin{align}
    \beta_\pm = i \omega\pm i\sqrt{\omega^2 - m^2} \mp \epsilon\,, 
\end{align}
the on-shell action diverges exponentially in the same way indicated in \eqref{eq:divergence}.

The steps of the holographic renormalization procedure lead to the Hamilton-Jacobi equation in the form of the Riccati equation \eqref{eq:GeneralRiccati}, which is the same as \eqref{eq:RiccatiVR},  except for the mass term, i.e. 
\begin{equation}
\partial_r f+\frac{d-1}{r}f + f^2  -\left(\partial_v^2 - \frac{1}{r^2}\Delta_{\hat{g}}\right)-m^2=0.
\end{equation}
The discussion of the solution of $f$ also proceeds unobstructed exactly as in the massless case and we can simply conclude that the renormalized on-shell action is \eqref{eq:SOnShellTRfreq}. The difference to the massless case is only captured by the different $\beta_\pm$ and hence the renormalized on-shell action and the canonical momentum read, in frequency space, explicitly as 
\begin{align}
\label{eq:SrenMassive}
    S_{\mathrm{os}}^{\mathrm{ren}} &= -\frac{i}{2 \pi} \int_{|\omega|\geq m} \dt \omega\, \dt \Omega  \, \tilde{\phi}^{(\mathrm{II})}(-\omega)\left(\sqrt{\omega^2 - m^2} -i\epsilon\right)\tilde{\phi}^{(\mathrm{I})}(\omega) \\ 
    \label{eq:massivePiRen}
    \Pi^{\mathrm{sub}} &= -\frac{i\sqrt{\hat{g}}}{2 \pi}r^{-\frac{d-1}{2}}e^{\beta_+ r } \left( \sqrt{\omega^2 - m^2}-i\epsilon\right)\tilde{\phi}^{\mathrm{(I)}}(\omega)\, . 
\end{align}
The development of a flat holography computational rule that accounts simultaneously for massless and massive fields is highly desired. Although massive and massless geodesics belong to different causal regions of the spacetime and its boundary, there is no a priori reason why massive and massless fields should be treated on different grounds. Also in AdS massive and massless geodesics behave starkly differently, with massive ones never reaching the conformal boundary. Yet, the GKPW dictionary treats them democratically.

\section{Interactions}
\label{sec:interactions}
To extend our construction to compute higher-point functions, we need to consider bulk interacting fields. In this section we propose a computation of a Carrollian three-point function at tree-level using a bulk self-interacting scalar field with potential
\begin{align}
    \label{eq:interactionTerm}
    V = -\frac{\lambda}{3} \Phi^3\,.  
\end{align} 
According to the general dictionary, we propose that
\begin{equation}
 \langle \mathcal{O}(X_1)\mathcal{O}(X_2)\mathcal{O}(X_3) \rangle= \left.\left(\frac{1}{\sqrt{\hat{g}}}\frac{\delta}{\delta \phi_s(X_2)}\frac{1}{\sqrt{\hat{g}}}\frac{\delta}{\delta \phi_s(X_3)}\right)\lim_{r_0\to\infty}\frac{1}{r^{\frac{d-1}{2}}e^{\beta_+ r}}\Pi^{\mathrm{sub}}(X_1)\right|_{\phi_s =0} \, ,
\end{equation}
where $\Pi^{\mathrm{sub}}$ is the subtracted canonical momentum of the interacting theory and $\phi_s$ is the appropriate source for the interacting theory. 

In section \ref{sec:holo_ren_int} we comment on the points that necessitate extreme care when setting up the Hamilton-Jacobi construction. In this regard, note that the counterterm ansatz cannot be the only term for non-quadratic theories. Assuming these issues can be resolved and that the identification of sources and expectation values with the same asymptotic field data established in the free case still holds, we proceed to subsection \ref{sec:propagators}, where we derive the propagators needed to set up a Witten–diagram–style computation in subsection \ref{sec:3pt}.  
\subsection{Set-up of the perturbative expansion}\label{sec:holo_ren_int} 
Following the strategy of \cite{vanRees:2011fr}, we solve the equations of motion
\begin{align}
    \left(\square_{g} - m^2 \right)\Phi = \lambda \Phi^2
\end{align}
perturbatively in $\lambda$. Therefore, we assume
\begin{align}
    \Phi = \sum_{k = 0}^{\infty} \Phi_k \lambda^k\,.
\end{align}
 
The equation of motion at order $\lambda^0$ is simply the equation of motion for a free scalar field $\Phi_0$. The equation of motion of $\Phi_1$ is instead sourced by $\Phi_0$, 
\begin{align}
\label{eq:interactions:Phi1equation}
    \left(\square - m^2\right) \Phi_1 = \Phi_0^2\, ,
\end{align}
and so on at higher orders. 
The solution of this inhomogeneous linear differential equation \eqref{eq:interactions:Phi1equation} is commonly obtained in terms of the bulk-to-bulk propagator $G(r, t, \Omega; r', t', \Omega')=:G(x; x')$ as
\begin{align}\label{eq:integralPhi1}
    \Phi_1(x) = \int \dt^{d+1}x'\, \sqrt{|g|} \, G(x; x') \, \Phi_0(x')^2\, . 
\end{align}
However, one must be extremely careful with the convergence of this integral. As we explicitly compute in section \ref{sec:bulkbulk}, the bulk-to-bulk propagator behaves as $G(x; x')\sim e^{\beta_+ r'}$ for $r'\to\infty$ with $r$ fixed and since $\Phi_0^2(x')\sim e^{2\beta_-r'}$, the integrand diverges as $e^{\beta_-r'}$ for any $r$ because $G$ is symmetric in its arguments. The 
naive large $r$ behavior of the integral does not agree with the large $r$ behavior of $G(x, x')$ as the integral receives corrections due to the divergences.\footnote{The same tricky situation is encountered for scalars in AdS when their characteristic $\Delta$ corresponds to an irrelevant conformal dimension of the dual CFT operator, as thoroughly discussed and resolved in \cite{vanRees:2011fr}.}

We can promptly and more conveniently obtain the general structure of the solution of \eqref{eq:interactions:Phi1equation} by simply noticing that it is the sum of a 
 free solution $\Phi_{h}$ behaving as 
 \begin{equation}
    \Phi_h=r^{-\frac{d-1}{2}}\left(e^{ \beta_+ r}(\bar{\phi}^{(\mathrm{I})}+\dots) +e^{ \beta_- r}(\bar{\phi}^{(\mathrm{II})}+\dots)\right)
 \end{equation}
 and a particular solution $\Phi_{p}$ of the form 
\begin{align}
 \Phi_p = r^{-\frac{d-1}{2}}\left(e^{2 \beta_- r} \left(A+\dots\right)+\left(B+ \dots\right)+e^{2 \beta_+ r}\left(C e^+\dots \right)\right)\,,  
\end{align}
with coefficients $A,B,C$ determined by $\Phi_0^2$. 

We can now make an educated guesses on the effect of holographic renormalization, once again inspired by the case of irrelevant scalars in AdS/CFT. Although $\Phi_1$ diverges\footnote{The terminology refers to the analytically continued fields.} faster than $\Phi_0$ due to $r^{-\frac{d-1}{2}}e^{2 \beta_- r}$, the orders of the two free asymptotic data of $\Phi_1$ are (not surprisingly) determined by the homogeneous solution. Thus at this perturbative order in $\lambda$ the holographic data of the free theory are modified as
\begin{equation}
\phi^{(\mathrm{I})}_0\to\phi^{(\mathrm{I})}_0+\bar{\phi}^{(\mathrm{I})}, \qquad \phi_0^{(\mathrm{II})}\to\phi^{(\mathrm{II})}_0+\bar{\phi}^{(\mathrm{II})}.
\end{equation}
It is natural to expect that the free orders of the asymptotic expansion of the bulk field retain their holographic interpretation,  captured by
\begin{align}
\langle \mathcal{O}(X)\rangle_{\phi_s} \sim \phi^{\mathrm{(I)}},
\end{align}
even in the interacting case. Recall that the source $\phi_s$ is related to $\phi^{\mathrm{(II)}}$ by \eqref{eq:defPhiSviaoperator}. This motivates the boundary condition
\begin{equation}
\bar{\phi}^{(\mathrm{II})} = 0 \, ,
\end{equation}
which implies that the source of the dual operator remains unchanged by interactions, as is commonly assumed in AdS/CFT.  However, a full confirmation of this assumption requires performing holographic renormalization. In fact, it is not automatically guaranteed that the counterterms for the free theory are enough to cancel the overleading divergences due to $\Phi_p$ at order $\lambda$. We postpone to future endeavour the task of constructing the counterterm order by order in $\lambda$, and we now assume that this is the case.\footnote{It would be interesting to assess how the holographic renormalization procedure for the interacting system handles the familiar IR divergences affecting massless theories.} 

With these assumptions, we can safely take the bulk–bulk propagator as the fundamental building block of Witten diagrams and proceed as discussed below, extracting the other relevant propagators through the limiting procedure applied to the free propagators.

\subsection{Propagators}\label{sec:propagators}
\subsubsection*{The bulk-to-bulk propagator}\label{sec:bulkbulk}
We define the bulk-to-bulk propagator as the solution of
\begin{equation}
\label{eq:defBulkBulkmain}
(\Box - m^2) G(x; x') = \frac{1}{\sqrt{|g|}} \delta^{d+1}(x - x')\, , 
\end{equation}
with appropriate boundary conditions corresponding to Feynman propagators. 
The equation \eqref{eq:defBulkBulkmain} has the solution
\begin{align}\label{eq:G}
\begin{split}
G(x; x') = & -\frac{i\pi}{2 }\int_{-\infty}^{\infty}  \frac{\dt \omega}{2\pi} e^{-i\omega(t - t')} \frac{1}{(r r')^{\frac{d-2}{2}}} \\& \times
\begin{cases}
J_{\nu} \left( \sqrt{\omega^2 - m^2} \, r' \right) H_{\nu}^{(1)} \left( \sqrt{\omega^2 - m^2} \, r \right)\delta(\Omega - \Omega') & \text{for } r > r' \\
J_{\nu} \left( \sqrt{\omega^2 - m^2} \, r \right) H_{\nu}^{(1)} \left( \sqrt{\omega^2 - m^2} \, r' \right)\delta(\Omega - \Omega') & \text{for } r' > r
\end{cases} \, .
\end{split}
\end{align}
The detailed calculation, including the $i\epsilon$ prescription, can be found in Appendix \ref{sec:bulkbulkDETAILS}. 
The $i\epsilon$ prescription as well as the radial step function ensures that the bulk-to-bulk propagator is fully normalizable. The step function is reminiscent of the bulk-to-bulk propagator in AdS spacetimes \cite{Mueck:1998wkz, Freedman:1998tz}.

\subsubsection*{The bulk-to-boundary propagator}
In order to compute three-point functions, bulk-to-boundary propagators are necessary building blocks as well. 
We define the bulk-to-boundary propagator $K(r, X; X')$, where $X'$ is a boundary and $(r, X)$ is a bulk point, as the limit 
\begin{align}
    K(r, X; X') = \lim_{r'\to \infty} \frac{1}{e^{\beta_+ r'} (r')^{-\frac
    {d-1}{2}}}  \left(\partial_{r'} -f(r')\right)G(r, X; r', X')\, . 
\end{align}
 This bulk-to-boundary propagator does not take into account the antipodal identification of the source introduced in \eqref{eq:defPhiSviaoperator}. 
It will be useful to write $f$ as a closed form expression in terms of the Hankel function (see footnote \ref{Hankelfootnote})
\begin{align}
    f = \partial_r \log \left(r^{-\frac{d-2}{2}}H^{(2)}_{\nu}\left(\sqrt{-\partial_t^2-m^2}r\right)\right)\, . 
\end{align}
Using the identities \eqref{eq:ToolForPropagotors_1}, \eqref{eq:ToolForPropagotors_2}, and \eqref{eq:crossproduct}, evaluating the asymptotics of Hankel function $H_{\nu}^{(1)}$ according to \eqref{eq:AsymptoticsHankelfunctions} and Fourier transforming via \eqref{eq:FTdelta}, we obtain

\begin{align}
\label{eq:bulkboundaryPropagator}
    K(r, X; X')  = \frac{\sqrt{2\pi } e^{-\frac{i\pi(2\nu+1)}{4}}}{r^{\frac{d-2}{2}}}J_\nu(\sqrt{-\partial_{t'}^2-m^2}r) \, \left(-\partial_{t'}^2-m^2 \right)^{1/4} \delta(t-t') \delta^{d-1}(\Omega-\Omega') \, ,
\end{align}
To check that $K$ is indeed the bulk-to-boundary propagator, we compute the regular solution to the equations of motions $\Phi_{\text{Reg}}(r',t', \Omega')$, defined by
\begin{equation}
\label{eq:bulkboundarypropagatorProperty}
\Phi_{\text{Reg}}(r,X) = \int \dt^d X' \, \sqrt{\hat{g}}\,  K(r, X; X') \,\phi^{\mathrm{(II)}}(X') \, .
\end{equation} 
Recall that the actual source $\phi_s$ is related to $\phi^{\mathrm{(II)}}$ via the antipodal map defined in \eqref{eq:defPhiSviaoperator}. 
The field $\Phi_{\text{Reg}}(r',t', \Omega')$ is the regular solution with source $\phi_s(t, \Omega)$. In particular, using \eqref{eq:defPhiSviaoperator}, we obtain
\begin{equation}
\Phi_{\text{Reg}}(r,t, \Omega) =  \sqrt{2\pi} \frac{e^{-i\pi\frac{2 \nu+1}{4}}}{( r)^{\frac{d-2}{2}}}J_\nu\left(\sqrt{-\partial_{t'}^2-m^2}r\right) \, \left(-\partial_{t}^2-m^2 \right)^{1/4} \phi^{\mathrm{(II)}}(t, \Omega) 
\end{equation}
which is in agreement with \eqref{eq:regularsolwithsource}. 
Later, it will be useful to work with the bulk-to-boundary propagator that is defined with respect to $\phi_s$ instead of $\phi^{\mathrm{(II)}}$. This is best expressed in frequency space. For the massless case we obtain
\begin{align}
    \label{eq:defKs}
    K_s(\omega, \Omega; \omega', \Omega') = \mathrm{sign}(\omega)^{\ell_j}\, r^{-\frac{d-2}{2}} \sqrt{2 \pi} e^{-i\pi \frac{2 \nu +1}{4}}J_{\nu}(|\omega|r)|\omega|^{\frac{1}{2}}\delta(\omega - \omega')\delta\left(\Omega - \Omega'\right)\, .  
\end{align}

\subsubsection*{The boundary-to-boundary propagator}
We can repeat the same operation with respect to the remaining bulk coordinate to obtain the boundary-to-boundary propagator
\begin{align}
    R(X; X')=  \lim_{r \to \infty} \frac{1}{e^{\beta_+ r}r^{-\frac{d-1}{2}} }\left(\partial_r - f\right) K(r, X; X'). 
\end{align}
Inserting the bulk-to-boundary propagator \eqref{eq:bulkboundaryPropagator}, we obtain
\small
\begin{align}
    \label{eq:BoundaryBoundarycomputed}
    \begin{split}
      R(X; X') =&\lim_{r \to \infty} \frac{r^{d-1}}{e^{\beta_+ r}r^{\frac{d-1}{2}}} \sqrt{\frac{\pi}{2}}\left(\frac{4i}{\pi  r}\right)e^{-\beta_- r}e^{ -i \pi \frac{2 \nu + 1}{2}}\left(-\partial_t^2 - m^2\right)^{\frac{1}{2}} \\ & \qquad\times \delta(t-t')\delta(\Omega - \Omega')r^{-\frac{d-2}{2}}\sqrt{\frac{\pi r}{2}} + O(r^{-1})\,  \\
 = &\,2 i \left(-\partial_t^2 - m^2\right)^{\frac{1}{2}} \delta(t-t')\delta(\Omega - \Omega')\,e^{ -  i \pi \frac{ 2\nu + 1}{2}}\, . 
\end{split}
\end{align}
\normalsize

\subsubsection*{Relation to the on-shell action}

The boundary-to-boundary propagator is related to the renormalized on-shell action by

\begin{align}
    -\frac{1}{2}\int \dt t\,  \dt \Omega \int \dt t'\, \dt \Omega' \, \phi_s\,(t, \Omega) \,R(t, \Omega; t', \Omega')\,\phi_s(t' , \Omega')\, .  
\end{align} 
Plugging in the expression for the boundary-to-boundary propagator \eqref{eq:BoundaryBoundarycomputed} and doing the primed integral yields $S^{\mathrm{ren}}_{\mathrm{os}}$ since
\begin{align}
\begin{split}
    S_{\text{os}}^{\mathrm{ren}}=&-i \int \dt t\, \int d\Omega  \, \phi_s(t, \Omega) \, \left(-\partial_t^2 - m^2\right)^{\frac{1}{2}} \,\phi_s (t, \Omega) \\ 
    =& -i\int \frac{d \omega}{ 2 \pi}  \, \int d\Omega  \,  \tilde{\phi}_s(\omega, \Omega) \,\left(\omega^2 - m^2\right)^{\frac{1}{2}} e^{ -  i \pi \frac{ 2\nu + 1}{2}} \,\tilde{\phi}_s(-\omega, \Omega)\, . 
   \end{split}
\end{align}
This coincides with the renormalized on-shell action for the massive field \eqref{eq:SrenMassive} and in the massless limit to \eqref{eq:SrenTRfrequencyspace} once $\phi^{\mathrm{(I)}}$ is related to $\phi^{\mathrm{(II)}}$ by \eqref{eq:regularityForSources} and further to $\phi_s$ by \eqref{eq:defPhiSviaoperator}. 

\subsection{Three point correlation function}\label{sec:3pt}

To exemplify the computation of the three-point function, we consider only the massless field. From the discussion in section \ref{sec:holo_ren_int}, the term of order $\lambda$ in the expansion of the canonical momentum $\Pi^{\mathrm{sub}}$ is proportional to $\left(\partial_r - f \right) \Phi_1$. This allows to obtain the three point function as
\begin{align}
\begin{split}
    \langle \mathcal{O}(X_1)\mathcal{O}(X_2)\mathcal{O}(X_3) \rangle_{\mathrm{tree}} &= \lim_{r \to \infty} \frac{1}{r^{-\frac{d-1}{2}} e^{\beta_+ r}} \left(\partial_r - f\right)\Phi_1\\ & = \int \dt^{d+1} x \,K_s(x; X_1)K_s(x; X_2)K_s(x; X_3)\, . 
    \label{eq:3ptFreqStartingpoint}
    \end{split}
\end{align}
Note that the bulk-to-boundary propagators $K_s$ here are defined in \eqref{eq:defKs} with respect to $\phi_s$ instead of $\phi^{\mathrm{(II)}}$. Hence the arguments of $\Phi_1$ in \eqref{eq:3ptFreqStartingpoint} have to be antipodally identified as described in \eqref{eq:defPhiSviaoperator}.
Inserting the bulk-to-boundary propagator and performing some manipulations shown Appendix \ref{sec:3ptFunctionFrequencyCalcs}, we arrive at 
\begin{align}
\label{eq:3pFunctionIntermediatemain}
\begin{split}
    \langle \mathcal{O}(\omega_1, \Omega_1)\mathcal{O}(\omega_2, \Omega_2)\mathcal{O}(\omega_3, \Omega_3) \rangle_{\mathrm{tree}} =& \prod^3_{j = 1}\left(\left(\frac{|\omega_j|}{2 \pi}\right)^{\frac{d-1}{2}}e^{\frac{-i\pi(d-1)}{4}}\right) \\ 
    & \times \delta \left(\hat{x}_1\omega_1+\hat{x}_2\omega_2+\hat{x}_3\omega_3 \right)\delta(\omega_1 + \omega_2 + \omega_3)\, , 
    \end{split}
\end{align}
where we expressed the angles $\Omega_j$ as unit-vectors  $\hat{x}_j$ on the sphere $\mathds{S}^{d-1}$ embedded in $\mathds{R}^{d}$. One can further use the coarea formula \cite{6fe0bdee-acb9-3218-a361-f3d025305446} (see also \cite{GiaquintaMathematicalAnalysis}) to make the collinear structure of the form $ \delta\left(\hat{x}_1 - \hat{x}_2\right)\delta\left(\hat{x}_1 - \hat{x}_3\right)$ manifest for the three point function. 
This feature is common to amplitudes for electric Carrollian field theories, e.g.  \cite{Bagchi:2023fbj}. 

Notice that \eqref{eq:3pFunctionIntermediatemain} agrees up to prefactors with the results of \cite{Kulkarni:2025qcx, Nguyen:2023miw}. For example, in  equation (3.8) of \cite{Kulkarni:2025qcx} the $\Delta_k$ has to be set equal to $\Delta$ in \eqref{eq:scalaingDimension} to obtain the same exponent in \eqref{eq:3pFunctionIntermediatemain}. There is, however, an interesting feature of our three-point function. Since the frequencies in \eqref{eq:3pFunctionIntermediatemain} range from $-\infty$ to $+ \infty$, and positive frequencies correspond to scattering data on $\scri^{-}$ and negative frequencies to data on  $\scri^{+}$, the three-point correlation function naturally comprises both, $1\to 2$ and $2 \to 1$ scattering configurations.

\section{Discussion}
\label{sec:Discussion}

In the search for holographic dualities for asymptotically flat quantum gravity theories that operate via mechanisms similar to those of the AdS/CFT correspondence, extending the GKPW prescription is a crucial step, as it does not depend on the specific AdS background in which it is applied. In particular, it first relates the one-point functions of dual CFT operators to renormalized bulk canonical momenta, and subsequently allows the extraction of two-point and higher-point correlation functions once the state -- namely, the bulk geometry -- is fully specified. This framework relies on an elegant interplay between asymptotic phase spaces and their holographic interpretation on the boundary.

In flat holography, this picture has long remained elusive due to the intricate structure of asymptotic boundaries and the widely held expectation that the dual non-gravitational theory -- Witten’s “structure X” -- must be capable of reconstructing bulk scattering processes.

In this paper, we  have demonstrated that it is indeed possible to closely follow the lessons of the GKPW prescription in flat holography, even when focusing on scattering boundary conditions, by appropriately adapting the framework of holographic renormalization.

The holographic renormalization procedure we have set up
has two main features. 
First, it manifestly associates a single bulk function to each putative boundary operator, while at the same time encoding the same  boundary conditions of the AFS generating functional: positive frequency data at early times (namely  on $\mathscr{I}^{-}$ for massless scattering) and negative frequency data at late times (e.g. $\mathscr{I}^{+}$).
Second,  the working mechanisms exemplified in this paper rely on general principles that admit systematic extensions. These, in turn, outline a path toward elevating the relation
\begin{equation}
\pi\sim\braket{\mathcal{O}}_{\phi_s}
\end{equation}
into a robust computational rule for flat holography, that goes beyond case-by-case extrapolations.

In these final pages, we offer additional discussion that was omitted from the technical development of the preceding sections, with the aim of highlighting key features and identifying directions for further exploration and extension.

A key observation in the technical development of the paper is related to the structure of the asymptotic solutions, which are of Thomé type, in both Lorentzian and Euclidean signature. For example, as discussed in section \ref{sec:SingularitiesOfDifferentialEquations}, in ingoing Eddington-Finkelstein coordinates,
one can rewrite the solution in frequency space as
\begin{equation}\label{eq:osc_power}
 \Phi \sim r^{-\frac{d-1}{2}}e^{-i\omega v}\left(\tilde{\phi}^{\mathrm{(p)}}(\omega)(1+\dots)+e^{2i\omega r}\tilde{\phi}^{\mathrm{(o)}}(\omega)(1+\dots)\right), 
\end{equation}
where $\tilde{\phi}^{\mathrm{(\mathrm{p})}}(\omega)$ is the data of the simple power-law behaviour and  $\Phi^{\mathrm{(o)}}$ the data of the the oscillating term. The asymptotics $\tilde{\phi}^{\mathrm{(p)}}$ and $\tilde{\phi}^{\mathrm{(o)}}$ are related to the functions $\phi^{\mathrm{(I)}}$ and $\phi^{\mathrm{(II)}}$ according to their frequencies as given in table \ref{tab:FrequenciesBranches}. 

It is common to assume that the asymptotic solution near either $\mathscr{I}^+$ or $\mathscr{I}^-$ is characterised by a power-law expansion, possibly with logarithmic terms, e.g. \cite{Bekaert:2022ipg,Briceno:2025ivl,Campoleoni:2025bhn}\footnote{The occurrence of polyhomogeneous expansions is a well-known phenomenon in general relativity (see e.g. \cite{Chrusciel:1993hx,ValienteKroon:2001pc,Capone:2021ouo,Kehrberger:2021uvf})  and has been investigated also in other gauge theories (see e.g. \cite{Campiglia:2019wxe,AtulBhatkar:2020hqz,AtulBhatkar:2021txo,Kim:2023qbl,Compere:2025tzr}). Typically, the logarithmic terms are mostly subleading in the large-$r$ expansion near  $\mathscr{I}^+$ and some of them are known to be associated to classical versions of soft theorems and the $1/u$ tails at null infinity \cite{Saha:2019tub,Geiller:2024ryw}. Logarithmic terms at the same large-$r$ order as the radiative term are also possible within general relativity in four and higher dimensions \cite{Chrusciel:1993hx,ValienteKroon:2001pc,Capone:2021ouo}, but their physical interpretation is less transparent or more exotic, e.g. \cite{Fursaev:2023oep}. 
Recently, \cite{Briceno:2025cdu} provided a neat example in four spacetime dimensions of how overleading logarithmic terms near $\mathscr{I}^+$ may appear also in a scalar theory due to $\log(v)$ behaviour of the advanced solution. A priori, the Fourier transform of $e^{-i\omega v} e^{2i\omega r}\tilde{\phi}^{\mathrm{(o)}}(\omega)$, e.g. a generic function $F(v-2r)$ could accommodate this behaviour if we intend the Fourier transform in a generalised sense. It would be interesting to explore the connections between the solutions of the massless PDE of \cite{Briceno:2025cdu} and our approach based on the radial ODE further. }
\begin{equation}\label{eq:power}
\Phi(v,r,\Omega)=r^{-\frac{d-1}{2}}\left(\phi^{(p)}(v,\Omega)+\dots\right).
\end{equation}
Formally, this misses the oscillating branch of \eqref{eq:osc_power}, which is necessary to prescribe the other half of the scattering data if we analyze the problem in a single coordinate system.

The Riemann-Lebesgue lemma \cite{JLBC1951} can be used to justify the solution ansatz \eqref{eq:power}. This is indeed consistent within  the studies on advanced or retarded ``sandwich" radiation, as originally explored by Bondi and collaborators for gravity. However, generic scattering configurations do not satisfy the conditions to apply the lemma, as the data are not $L^1$-integrable (e.g. the familiar $1/u$ tail at null infinity). For example, \cite{Kraus:2025wgi} noted that variational derivatives with respect to what we call $\phi_s$ cannot be defined unless $\phi_s$ does not belong to the space of $L^1$-integrable functions. 
To recover a solution appropriate to scattering (at least classically) one is then required to perform the analysis also in retarded coordinates, asymptotically match the expansions and eventually impose regularity conditions in the interior to have well-defined solutions. This approach is common in asymptotic studies of gravitational radiation (where there are good reasons to not impose bulk conditions), but it is surely not necessary for linear fields in fixed backgrounds.  

Once the importance of taking the full Thomé solution into account is recognised, we run in the problem that the on-shell action is not defined in the large $r$ limit.\footnote{In contrast, comparing with the discussion in section \ref{sec:need_holo}, the solution \eqref{eq:power} gives an automatically vanishing on-shell action up to the terms we also discarded.  \label{footnote:vanishingaction}} To deal with this, we first analytically continue the frequencies in accordance with the Feynman contour and then renormalize the divergent on-shell action. 

Our holographic renormalization approach follows the Hamilton-Jacobi procedure, but it necessarily departs from its standard usage within the AdS/CFT correspondence. 

For simplicity we used coordinates manifestly adapted to a  foliation  of the spacetime in terms of timelike surfaces. Ideally, the counterterm should be manifestly covariant on the surface. However, as we have mentioned in section \ref{sec:Holographicrenormalizationformasslessfields}, we refrain from retaining full covariance since we eventually take the large $r$ limit (the same consideration would apply to a conformally compactified version of the computation). Apart from this, the situation is pretty much in parallel with AdS/CFT. To construct a procedure that embeds covariance in a better way, it would be interesting to redo the analysis with an explicit geometric treatment of the regulator surfaces and their limit. More insightfully, working from the outset with a double–null foliation would allow one to maintain boundary symmetries in a manifestly covariant manner. Particularly convenient are the so-called flat-null coordinates, where the metric at the intersections of the null surfaces is planar (see for example Appendix A of \cite{He:2019jjk}). The Hamilton-Jacobi equation becomes, though, that for a constrained system and thus requires more care. 

Another feature of standard AdS/CFT holographic renormalization that we needed to give up is locality.  In (asymptotically) flat spacetime, it is widely known that the status of locality of the counterterms is problematic. This issue has long been associated with the fact that subleading terms in asymptotic solutions are not algebraically determined by the leading data \cite{deHaro:2000wj}. In this paper we have discussed extensively another form of violation of locality, related to the fact that divergences are exponential and require an infinite number of counterterms.
These are however all captured by one single expression.

The question arises naturally: why should one insist on the locality of counterterms in flat holography? In the context of AdS/CFT, the standard justification is that the dual conformal field theory is local, and its ultraviolet divergences are removed by local counterterms. This is often understood as a non-trivial check of the the IR/UV connection. However, there exist at least one known example in which strictly local counterterms are insufficient to renormalize all divergences. This is the case of a self-interacting irrelevant scalar field \cite{vanRees:2011fr,vanRees:2011ir}.

In our work, insisting on locality of counterterms does not appear to be justified because we seek a holographic reconstruction of the bulk from   scattering data and this forces us to deal with the Thomé solutions. This aligns with, and further supports, recent observations made in \cite{Cotler:2025npu}.

We have shown that the holographic renormalization procedure we propose yields for free massless bulk fields the Carrollian two-point function \eqref{eq:Carollian2ptfunctionTR}. The counterterms we construct to reach this result contain an ambiguity, indicated by the function $C$ in \eqref{eq:ratio}, that we have set to zero motivated by the request that the source and the expectation value form a well-defined canonical pair in the asymptotic phase space. 

We also suggest that this holographic renormalization approach applies to massive bulk fields, although the holographic interpretation of the massive sector requires further investigation. The resulting uniformity of the putative computational rule underlying the holographic dictionary is, in our view, a particularly desirable feature of the framework.

As a further outlook, we have initiated a study of interacting bulk scalars within this formalism. As a first step, we considered a $\lambda\Phi^3$ theory, and exemplified how to obtain a three-point function using Witten-diagram techniques. In this analysis, we assumed that the bulk data identifying the dual source and expectation value are the same as in the free theory, in line with standard treatments in AdS/CFT. In section \ref{sec:holo_ren_int} we offer cautionary remarks about this assumption, in parallel to analogous studies of irrelevant scalar operators in AdS/CFT \cite{vanRees:2011fr,vanRees:2011ir}. We believe that a systematic treatment of these points constitutes an important direction for future work. It is particularly interesting to assess the imprint of the familiar IR divergences in massless scattering amplitudes on this holographic renormalization procedure. A fully consistent approach is expected to deliver IR-finite results and observables.

Both the two and three point Carrollian functions that we derive match existing results in literature \cite{Kulkarni:2025qcx, Nguyen:2023miw, Donnay:2022wvx, Mason:2023mti, Kraus:2024gso}, but there are two interesting differences. The first concerns the identification of the scaling dimension of the dual Carrollian operator. The identification that follows naturally from our holographic renormalization does not correspond to the simple leading power of the asymptotic expansion, e.g. the scaling dimension of the bulk scalar. In fact, in the GKPW-like dictionary the asymptotic behaviour of the field constitutes the source of the dual operator, not the operator itself. The second concerns the structure of the correlators with more than two insertions. As seen at the end of the previous section, an $N$-point Carrollian correlation function is made of summands that corresponds to all possible kinematical configurations, i.e. scattering amplitudes with $m$ ingoing particles and $N -m$ outgoing particles with $m = 1, \dots, N-1$. 
\\
\\
We conclude by pointing towards possible generalisations. Extending our analysis to other free fields of higher spin, including linearized gravity on a Minkowski background, appears to be straightforward. In these cases we also expect our approach to yield a different identification for the conformal dimensions of the dual conformal Carrollian operators compared to the conformal Carrollian representations built in  e.g. \cite{Nguyen:2023vfz}.

It is also noteworthy that our formalism can be adapted to curved backgrounds. Certain backgrounds, such as shock waves and cosmic strings, allow for exact solutions of the wave equations \cite{Moreira:1995he}. We expect that our holographic renormalization approach can be applied to these cases with minimal modifications. More intriguingly, when applied to black hole backgrounds, our approach may prove more convenient than the use of the AFS path integral with time foliation. In fact, within our approach, we would only need to consider asymptotic solutions. Moreover, we only have to impose boundary conditions at infinity, in line with Lorentzian AdS/CFT literature \cite{Skenderis:2008dg}, while in  the AFS path integral with time foliation we may have to impose boundary conditions at the horizon in addition to those at infinity. 

Clearly, extending our proposal to dynamical, nonlinear gravity is another natural direction. While working in standard Bondi coordinates may be technically simpler, as the general structure of the solutions is already explicitly known to some orders, it would be both interesting and advantageous to formulate the analysis in double null coordinates.

We emphasize that our holographic renormalization has been developed in the context of scattering boundary conditions. As such, direct comparison with the recent works \cite{Hartong:2025jpp,Campoleoni:2025bhn} or with the earlier analysis of  \cite{Bagchi:2015wna} is not possible beyond the comments we gave around equation \eqref{eq:power} and footnote \ref{footnote:vanishingaction}, because they focus on asymptotic expansions of (retarded) fields and gravity at $\mathscr{I}^+$. Nonetheless, if one wishes to consider these scenarios, the Hamilton-Jacobi procedure may still provide novel insights. In general, the Hamilton–Jacobi counterterm equation with different boundary conditions (for instance, adopting an $i\epsilon$ prescription that does not correspond to the Feynman choice) would be different. It is tantalizing to ask whether such an analysis could offer insights into new observables in flat holography.

Another feasible study is the comparison of the Hamilton-Jacobi renormalization procedure to actions and asymptotic charges at spacelike and null infinity, revisiting with this technique the vast body of work that we have mentioned in the introduction, as well as \cite{Laddha:2020kvp,Laddha:2022nmj,Compere:2023qoa}. 

In conclusion, the Hamilton-Jacobi holographic renormalization procedure appropriately adapted to flat spacetime provides both a valuable approach to construct a GKPW dictionary within flat holography and a very versatile methodology to explore covariant phase spaces.

\section*{Acknowledgements}
We thank Glenn Barnich, Dominik Neuenfeld, Ana-Maria Raclariu and Romain Ruzziconi for related discussions. FC is grateful to Miguel Campiglia and the University of Montevideo for their hospitality during various stages of this project, as well as to Guillermo Silva and his students for stimulating discussions on related topics during these stays. MA, FC and CS thank the Erwin-Schrödinger Institute (ESI) for the hospitality in April 2024 during the program ``Carrollian physics and holography". FC and CS are supported by the Deutsche Forschungsgemeinschaft (DFG) under
Grant No 406116891 within the Research Training Group RTG 2522/1.

\appendix
\section{Theorems on  ODEs}\label{app.ode}

In this appendix, we review the properties of asymptotic solutions of differential equation. 
We mostly follow \cite{Slavyanov}. Given a linear, second-order differential equation, possibly in $\mathbb{C}$, of the form
\begin{equation}\label{eq:ODE}
 P(x)y''+Q(x)y'+R(x)y=0\,,  
\end{equation}
we have the following:

\paragraph{Definition (singular point):} A point $x_0$ is a singular point if $P(x_0)=0$.

\paragraph{Definition (Fuchsian singular point):} The point $x_0$ is a Fuchsian point if 
\begin{equation}
\lim_{x\to x_0}(x-x_0)\frac{Q(x)}{P(x)} \quad \text{and} \quad \lim_{x\to x_0}(x-x_0)^2\frac{R(x)}{P(x)} \quad \text{are both finite.}
\end{equation}
Rewriting the equation as $y''+p(x)y'+q(x)y=0$, the conditions are that  $\lim\limits_{x\to x_0}(x-x_0)p(x)$ and $\lim\limits_{x\to x_0}(x-x_0)^2q(x)$ are both finite.

\paragraph{Definition (regular singular point):} The point $x_0$ is a regular singular point if the two linearly independent particular solutions of \eqref{eq:ODE} are functions of finite order at $x_0$, that is, both of them are functions $f(x)$ such that 
\begin{equation}
\lim_{x\to x_0}(x-x_0)^\rho f(x)=0, \quad \text{for some real number } \rho.
\end{equation}
A function that satisfies the above condition is said to be of finite order at $x_0$.
\paragraph{Definition (irregular singular point):} The point $x_0$ is an irregular singular point if it is a singular point and does not satisfy the former definition. 

\paragraph{Definition (s-rank):} The s-rank $R(x_0)$ of an irregular singular point is defined as $R(x_0)=\max(K_1(x_0),\frac{K_2(x_0)}{2})$, where $K_1(x_0)$, $K_2(x_0)$ are respectively the multiplicities of the poles of $p(x)$ and $q(x)$ at $x_0$. This definition makes the $s$-rank of a regular singular point equal to one. 

\paragraph{Definition (ramified/unramified singularity):} Irregular singular points with half-integer $s$-rank are called ramified. Irregular singular points with integer $s$-rank are called unramified. \\

\noindent With obvious modifications, all these definitions apply also to the point $x_0=\infty$. The notion of singular point captures the fact that the Cauchy problem cannot be solved around such an $x_0$. It might seem obvious that the definition of regular singular point is equivalent to the definition of Fuchsian singularity of the field equation. However, this is the content of a theorem which is not true in general for systems of equations.\footnote{The Bessel equation $x^2y''+xy'+(x^2-\alpha^2)y=0$ has a regular singular point at $x=0$ (which makes the parallel with the scalar field equation in AdS written in the coordinate $z$, and in fact as we said the closed-form solutions are Bessel functions of order $\alpha$). The point $x=\infty$ in the Bessel equation is an irregular singular point. The Euler equation is an example of an equation where all the singular points are Fuchsian. } The form of the solutions around singular points depend, as reported below, on the above definitions. 

\subsection{Solutions around regular singular points: Frobenius solutions}
Around a regular singular point one of the two independent solutions is always of the form of an analytic power series
   \begin{equation}
   	y_1(x)=|x-x_0|^\alpha\sum_{n=0}^\infty a_n(x-x_0)^n.
   \end{equation}
   The other solution, $y_2(x)$, varies according to the cases classified by the roots of the indicial equation
   \begin{equation}
   	\alpha(\alpha-1)+p_0\alpha+q_0=0\,,
   \end{equation}
   which we will call $\alpha_1$ and $\alpha_2$. We have the following cases:
   \begin{itemize}
       \item $\alpha_1\neq \alpha_2$ and $\alpha_1-\alpha_2 \notin \mathbb{Z}$: the second solution is of the power-law type 
   \begin{equation}
   	y_2(x)=|x-x_0|^{\alpha_2}\sum_{n=0}^\infty b_n(x-x_0)^n.
   \end{equation}
       \item $\alpha_1=\alpha_2$:  the second solution is
   \begin{equation}
   	y_2(x)=y_1(x)\log (x-x_0)+|x-x_0|^{\alpha_1}\sum_{n=0}^\infty b_n(x-x_0)^n.
   \end{equation}
       \item $\alpha_1>\alpha_2$ and $\alpha_1-\alpha_2\in \mathbb{N}$  (applies to the real part if they are complex): the second solution is
       \begin{equation}
   	y_2(x)=c\, y_1(x)\log (x-x_0)+|x-x_0|^{\alpha_2}\sum_{n=0}^\infty b_n(x-x_0)^n,
   \end{equation}
   where $c$ may as well be zero.
   \end{itemize}
   All these cases are realised by the scalar field in AdS according to the relative values of $\Delta_+$ and $\Delta_-$. 

Both of these solutions are called Frobenius solutions in the mathematical literature. 

\subsection{Solutions around irregular singular points: Thomé solutions}\label{app.solutionirreg}
Solutions cannot be of Frobenius form around an irregular singular point. In this case, the solutions in the form of asymptotic series are attributed to Thomé. If the order of the singularity is integer (unramified), the two independent solutions are both of the form 
\begin{equation}\label{eq:generalThome1}
y(x)=(x-x_0)^{\alpha} \, e^{\sum\limits_{k=1}^{R(x_0)-1}\frac{\beta_k}{(-k)}(x-x_0)^{-k}}\sum_{k=0}^{\infty} c_k(x-x_0)^k\, , 
\end{equation}
with coefficients $\alpha,\beta_k,c_k$ differing among the two solutions. If the order of the singularity is half-integer (ramified), the 
the two independent solutions are both of the form 
\begin{equation}\label{eq:generalThome2}
y(x)=(x-x_0)^{\alpha} \, e^{\sum\limits_{k=1/2}^{R(x_0)-1}\frac{\beta_k}{(-k)}(x-x_0)^{-k}}\sum_{k=0}^{\infty} c_k(x-x_0)^{k/2}\, , 
\end{equation}
with the sum in the exponent running over both integer and half-integer $k$, and coefficients $\alpha,\beta_k,c_k$ differing among the two solutions. The coefficient $\alpha$ is called characteristic multiplier of the Thomé solution and the coefficients $\beta_k$ are called characteristic exponents of the second kind of order $k$. These can be computed with recursion formulae. Analogous solutions exist for the irregular singular point at infinity (substitute $(\alpha,k)\to(-\alpha,-k)$ while keeping the sums over positive numbers).
It is not excluded in general that the sum contains logarithmic terms as well. However, such terms do not play a role in our analysis.

\section{Relations among different solutions of the scalar field}
\label{sec:RelationsAmongSolutions}
\subsection{Relation between closed form solution and asymptotic solution}
In this subsection we explicitly derive the relation between $\phi^{(\mathrm{I})}(t, \Omega)$ and $\phi^{(\mathrm{II})}(t, \Omega)$ of the asymptotic solution \eqref{eq:TRasymptoticFullFieldRepackaged}, and $\phi_1(t,\Omega)$ and $\phi_2(t,\Omega)$ appearing in the closed-form solution \eqref{eq:Formal_massive_tr_solution}. 
In order to do so we recall the asymptotic series of the Hankel functions. At large argument, the Hankel functions may be expanded as (according to equations 10.17.5 and 10.17.6 in \cite{NIST:DLMF})
\begin{align}
\label{eq:AsymptoticsHankelfunctions}
    H^{(1)}_{\nu}(x) &\sim \sqrt{\frac{2}{\pi x}}e^{i(x - \frac{1}{2}\nu \pi - \frac{1}{4}\pi)}\left(1 + O(x^{-1})\right)\,, \\
    H^{(2)}_{\nu}(x)& \sim \sqrt{\frac{2}{\pi x}}e^{-i(x - \frac{1}{2}\nu \pi - \frac{1}{4}\pi)}\left(1 + O(x^{-1})\right). 
\end{align}
This expansion allows us to compare the asymptotic expansion \eqref{eq:TRasymptoticFullFieldRepackaged} with the massless limit of \eqref{eq:Formal_massive_tr_solution}. We can relate the free functions in the following way
\begin{align}
\label{eq:TranslatingAsympVSExact}
    \phi^{(\mathrm{I})}(t, \Omega) &= \sqrt{\frac{2}{\pi }} \, e^{-\frac{i}{2}\left(\nu \pi + \frac{\pi}{2}\right)} \, \sqrt{-\partial_t^2-m^2}^{-\frac{1}{2}} \, \phi_1(t,\Omega) \, ,\\
    \phi^{(\mathrm{II})}(t, \Omega) &= \sqrt{\frac{2}{\pi }} \, e^{\frac{i}{2}\left(\nu \pi + \frac{\pi}{2}\right)} \, \sqrt{-\partial_t^2-m^2}^{-\frac{1}{2}} \, \phi_2(t,\Omega)\,, 
\end{align}
with $\nu$ given by \eqref{eq:nuDef}. In this very formal language, care must be taken to ensure that the operator $\sqrt{-\partial_t^2-m^2}^{-\frac{1}{2}}$ is well-defined. This operator is more rigorously defined in terms of its Fourier transformation, for which we find
\begin{align}
    \tilde{\phi}^{\mathrm{(I)}}_{0}(\omega, \ell, I) &= \sqrt{\frac{2}{\pi}} e^{-\frac{i}{2}(\nu \pi + \frac{\pi}{2})} \sqrt{\omega^2 -m^2}^{-\frac{1}{2}} \,\tilde{\phi}_{1}( \omega, \ell, I)\, , \\
    \tilde{\phi}^{\mathrm{(II)}}_{0}(\omega, \ell, I) &= \sqrt{\frac{2}{\pi}} e^{\frac{i}{2}(\nu \pi + \frac{\pi}{2})} \sqrt{\omega^2 -m^2}^{-\frac{1}{2}} \,\tilde{\phi}_{2}(\omega, \ell, I)\, . 
\end{align}
Here, $\tilde{\phi}_{1}$ and $\tilde{\phi}_{2}$ were defined in \eqref{eq:solutiontr} whereas 
$\tilde{\phi}^{\mathrm{(I)}}$ and $\tilde{\phi}^{\mathrm{(II)}}$ were defined in \eqref{eq:TRasymptoticFullField}.

\subsection{Remark about reality of the field}
As we want to consider real scalar fields, we have to ensure that the field is in fact real. 
Using equation 10.11.9 of \cite{NIST:DLMF}
\begin{align}
    H^{(2)}_{\nu}(z)^* = H^{(1)}_{\nu}(z^*)\, ,  \qquad H^{(1)}_{\nu}(z)^* = H^{(2)}_{\nu}(z^*) \qquad \text{for\, }\nu \in \mathds{R}\,,
\end{align}
and the fact that we can choose the spherical harmonics to be real, 
the reality condition constrains the complex number $\tilde{\phi}_{1, 2}$ that were defined in \eqref{eq:solutiontr} as
\begin{align}
    \tilde{\phi}_{1}(\omega ,\ell, I) = \tilde{\phi}_{2}(- \omega, \ell, I)^{*}\,, \qquad \tilde{\phi}_{2}( \omega, \ell, I) = \tilde{\phi}_{1}( -\omega, \ell, I)^{*}.
\end{align}
In this argument, we have implicitly considered $\omega$ with $|\omega| \geq m$ (see also appendix \ref{sec:planewaveExpansion}). 
Note that as we use real spherical harmonics, the sign in front of the $I$ does not change. 

By using \eqref{eq:AsymptoticsHankelfunctions}, we can conclude 
\begin{align}
\label{eq:relatitySourcesFourier}
\begin{split}
    \tilde{\phi}^{\mathrm{(I)}}(\omega, \ell, I)^{*} &=  \tilde{\phi}^{\mathrm{(II)}}(-\omega, \ell, I)\,,\\ 
    \tilde{\phi}^{\mathrm{(II)}}( \omega, \ell, I)^{*} &= \tilde{\phi}^{\mathrm{(I)}}(-\omega, \ell, I)\, .  
    \end{split}
\end{align}
The relation can also be directly deduced from \eqref{eq:TRasymptoticFullField}. 
In position space the relation reads
\begin{align}
    \label{eq:relatitySourcesreal}\begin{split}\phi^{\mathrm{(II)}}(\omega, \ell, I)^{*} &= \phi^{\mathrm{(I)}}(-\omega, \ell, I)\,,\\
    \phi^{\mathrm{(I)}}(\omega, \ell, I)^{*} &= \phi^{\mathrm{(II)}}(-\omega, \ell, I)\, . 
    \end{split}
\end{align}

\subsection{Relation to plane waves}
\label{sec:planewaveExpansion}
In this Appendix, we discuss the relation of the solutions for a massive real scalar field in $(t,r)$-coordinates to the plane wave solution of Cartesian coordinates $(t,\vec{x})$ with $\vec{x} \in \mathbb{R}^{d}.$ Its mode decomposition is given by
\begin{equation}
\Phi(t,\vec{x}) = \int \frac{\dt^d \vec{k}}{(2\pi)^d \, 2 \omega_{\vec{k}}} \, \left( a(\vec{k}) \, e^ {-i \omega_{\vec{k}} t + i \vec{k} \cdot \vec{x}} + a(\vec{k})^ {*} \, e^{i \omega_{\vec{k}} t - i \vec{k} \cdot \vec{x}}  \right),\label{eq:carteseansolutionddim}
\end{equation}
where we defined $\omega_{\vec{k}}  = \sqrt{k^2 + m^2}$ with $k^2 = \vec{k}^2$. Note that this implies that $\omega\geq m$. In the following, we use the identity \cite{moloi2022sphericalbesselfunctions}\,\footnote{This relation is usually stated with either Legendre polynomials or complex valued spherical harmonics; for the relation of Legendre polynomials in terms of real valued spherical harmonics see \cite{Frye:2012jj}. }
\begin{equation}
    e^{i \vec{k} \cdot \vec{x}} = (2\pi)^{d/2} \, (k r)^{1-\frac{d}{2}} \, \sum_{\ell=0}^\infty \sum_I i^\ell \, J_{\ell - 1 + \frac{d}{2}}(k r) \, Y_{\ell, I}(\hat{\vec{k}}) \, Y_{\ell, I}(\hat{x}) \, ,\label{eq:Bessel-sphericalharmonicsddim}
\end{equation}
where $r = \sqrt{\vec{x}^2}$ and $k = \sqrt{\vec{k}^2}$. Moreover, the unit vectors $\hat{\vec{k}}$ and  $\hat{x}$ are defined by $\hat{\vec{k}} = \vec{k} / k$ and $\hat{x} = \vec{x} / r,$ respectively. There is a noteworthy difference between the expansion for the plane wave in even- and odd-dimensional Minkowski-spacetimes. For even dimensional Minkowski spacetimes $d$ is odd, and hence the order of the Bessel function is half-integer and can be rewritten in terms of spherical Bessel functions, while in case of odd-dimensional Minkowski spacetime (where $d$ is even) the order of the Bessel function is integer.

Using the plane wave expansion \eqref{eq:Bessel-sphericalharmonicsddim} the mode expansion of $\Phi(t, \vec{x})$ may be rewritten as
\begin{align}
     \Phi(t, \vec{x})
    &=\frac{1}{r^{\frac{\dt}{2}-1}} \, \sum_{\ell, I} \int\limits_0^{\infty} \! \dt k\, J_{\ell-1+\frac{d}{2}} (k r) \, \left( e^{-i\omega_{\Vec{k}}t} \, \Tilde{a}_{\ell, I}(k) + e^{i\omega_{\Vec{k}}t} \, \Tilde{a}_{\ell, I}(k)^{*}  \right) \, Y_{\ell, I}(\hat{x})  \,,\label{eq:modedecompBessel}
\end{align}
with 
\begin{align}
\begin{split}
\Tilde{a}_{\ell, I}(k) &= \frac{k^{d/2}}{(2\pi)^{d/2} \, 2 \omega_{\vec{k}}}  \int \dt^{d-1}\hat{\Vec{k}}\, Y_{\ell, I}(\hat{\vec{k}}) \,  i^\ell a(\Vec{k})\,, \\
\Tilde{a}_{\ell, I}(k)^{*} &= \frac{k^{d/2}}{(2\pi)^{d/2} \, 2 \omega_{\vec{k}}}  \int \dt^{d-1}\hat{\Vec{k}}\, Y_{\ell, I}(\hat{\vec{k}}) \,  (-i)^\ell a(\Vec{k})^{*}\, . 
\end{split}
\end{align}
To demonstrate the equivalence between equation \eqref{eq:modedecompBessel} and the mode expansion presented in \eqref{eq:solutiongeneraltr}, we express the integral over \( k \in [0, \infty) \) in equation \eqref{eq:modedecompBessel} in terms of an integral over the frequency $\omega_{\vec{k}} = \sqrt{k^2 + m^ 2}$ (from now on we will drop the label $\vec{k}$ and just refer to the frequency $\omega$). In order to incorporate both dependencies $e^ {\pm i \omega t}$ in \eqref{eq:modedecompBessel} we extend the frequencies also to negative values. We obtain
\begin{align}
     \Phi(t, \vec{x})
    &=\frac{1}{r^{\frac{\dt}{2}-1}} \, \sum_{\ell, I} \int\limits_{|\omega|\geq m} \! \dt \omega\,  b_{\ell, I}(\omega) \, J_{\ell-1+\frac{d}{2}} (\sqrt{\omega^ 2- m^2} r) \, e^{-i\omega t}  \, Y_{\ell, I}(\hat{x}) \,, \label{eq:modedecompBessel2}
\end{align}
with
\begin{equation}
    b_{\ell, I}(\omega) = \frac{\sqrt{\omega^2 -m^2}^{\frac{d}{2}-1}}{(2 \pi)^{\frac{d}{2}}2} \int \dt^{d-1}  \hat{\vec{k}}\, Y_{\ell, I}(\hat{\vec{k}}) i^{\ell} a(\vec{k})\,,
\end{equation}
for $\omega > m$ and $b_{\ell, I}(\omega) = \left( b_{\ell, I}(-\omega) \right)^\star$ for negative values of $\omega$. The Bessel function $J_{\nu}(x)$ can be written in terms of Hankel functions as $J_{\nu}(x)=\frac{1}{2}\left(H^{(1)}_{\nu}(x)+H^{(2)}_{\nu}(x)\right)$. We expand \eqref{eq:modedecompBessel2} around $r\to\infty$ using \eqref{eq:AsymptoticsHankelfunctions} and obtain the leading behavior of the field
\begin{align}\label{eq:planewaveexpansionasymptotics}
\Phi(t,\vec{x})\sim \frac{1}{r^{\frac{d-1}{2}}}\sum_{\ell,I}\int\limits_{\omega\geq m} \! \dt \omega\,  \frac{Y_{\ell, I}(\vec{\hat{x}}) }{\sqrt{2 \pi k}} &\left[\left(b_{\ell , I}(\omega)e^{-i( \omega t-k r)}+ b_{\ell, I}(\omega)^* e^{i (\omega t+kr)}\right)e^{i (-\frac{1}{2} \nu \pi - \frac{1}{4} \pi)}\right.\nonumber \\
&+\left.\left(b_{\ell, I}(\omega) e^{-i( \omega t + k r)} + b_{\ell, I}(\omega )^* e^{ i (\omega t- kr)} \right)e^{i \left(\frac{1}{2} \pi \nu +\frac{1}{4} \pi\right)} \right]\,, 
\end{align}
where $k = \sqrt{\omega^2 - m^2}$  and $\nu = \ell + \frac{d}{2}-1$. In the massless case, $k=|\omega|$. This implies  $k=\omega \ge0$  in the integrand of \eqref{eq:planewaveexpansionasymptotics}.
Hence the leading term in the expansion of the massless field, expressed in terms of $v$ and $u$, can be written as
\begin{align}\label{eq:planewaveexpansionasymptotics_massless}
\Phi(t,\vec{x})=\frac{1}{r^{\frac{d-1}{2}}}\sum_{\ell,I}\int\limits_{\omega\geq 0} \! \dt \omega\,  \frac{Y_{\ell, I}(\vec{\hat{x}}) }{\sqrt{2 \pi \omega}} &\left[\left(b_{\ell , I}(\omega)e^{-i \omega u}+ b_{\ell, I}(\omega)^* e^{i\omega v}\right)e^{i (-\frac{1}{2} \nu \pi - \frac{1}{4} \pi)}\right.\nonumber \\
&+\left.\left(b_{\ell, I}(\omega) e^{-i\omega v} + b_{\ell, I}(\omega )^* e^{ i \omega u} \right)e^{-i \left(-\frac{1}{2} \pi \nu - \frac{1}{4} \pi\right)}\right]\,. 
\end{align}
Taking the limit $r\to\infty$ while keeping $v$ fixed, applying the Riemann-Lebesgue theorem, we obtain the field at $\mathscr{I}^-$
\begin{align}
 \Phi(t,\vec{x})=\frac{1}{r^{\frac{d-1}{2}}}\sum_{\ell,I}\int\limits_{\omega\geq 0} \! \dt \omega\,  \frac{Y_{\ell, I}(\hat{x}) }{\sqrt{2 \pi \omega}} &\left[b_{\ell, I}(\omega) e^{-i\omega v}e^{-i \left(-\frac{1}{2} \pi \nu - \frac{1}{4} \pi\right)} +b_{\ell, I}(\omega)^* e^{i\omega v}e^{i (-\frac{1}{2} \nu \pi - \frac{1}{4} \pi)}\right]\,,    
\end{align}
similarly, keeping $u$ fixed we get the field at $\mathscr{I}^+$ 
\begin{align}
\label{eq:planewaveexpansionasymptoticsII}
 \Phi(t,\vec{x})=\frac{1}{r^{\frac{d-1}{2}}}\sum_{\ell,I}\int\limits_{\omega\geq 0} \! \dt \omega\,  \frac{Y_{\ell, I}(\hat{x}) }{\sqrt{2 \pi \omega}} \left[b_{\ell, I}(\omega) e^{-i\omega u}e^{i (-\frac{1}{2} \nu \pi - \frac{1}{4} \pi)}+b_{\ell, I}(\omega)^* e^{i\omega u}e^{-i \left(-\frac{1}{2} \pi \nu - \frac{1}{4} \pi\right)} \right]\,.     
\end{align}
For the massive case, the analogue of \eqref{eq:planewaveexpansionasymptotics_massless} is
\small
\begin{align}
    \Phi = \frac{1}{r^{\frac{d-1}{2}}}\sum_{\ell, I} \int_{|\omega| \geq m} \frac{ \dt \omega\,  Y_{\ell, I}(\hat{x})}{\sqrt{2 \pi \sqrt{\omega^2 - m^2}}}&\left[\left(b_{\ell, I}(\omega)e^{-i \omega u} + b_{\ell, I}(\omega)^* e^{i\omega v}\right) e^{i \left(-\frac{1}{2} \pi \nu  - \frac{1}{4} \pi\right)}e^{i\left(\sqrt{\omega^2 -m^2}-\omega\right)r} \right. \nonumber \\
    & \hspace{-1cm}+ \left.\left(b_{\ell, I}(\omega)e^{-i \omega v} + b_{\ell, I}(\omega)^* e^{i\omega u}\right) e^{-i \left(-\frac{1}{2} \pi \nu  - \frac{1}{4} \pi\right)}e^{-i\left(\sqrt{\omega^2 -m^2}-\omega\right)r} \right]\, . 
\end{align}
\normalsize

\section{Useful identities}
\label{sec:Usefulidentities}
\subsection{Variational identities} 
Given the metric 
\begin{align}
    \dt s^2 &=  \dt r^2 - \dt t^2 + \gamma_{ij}(r,\vartheta)\dt \vartheta^i \dt \vartheta^{j} = \dt r^2 - \dt t^2 + r^2\hat{g}_{ij}(\vartheta)\dt \vartheta^i \dt \vartheta^{j}\, , 
\end{align}
with  $  R_{ij}[\gamma] = \frac{d-2}{r^2}\gamma_{ij}$, we have the following
\begin{align}\label{eq:identities1}
 r = \sqrt{\frac{(d-1)(d-2)}{R[\gamma]}} ,
\end{align}
where $K$ is the extrinsic scalar curvature of the $r=const$ surface. 
We also have the following variational identities
\begin{align}\label{eq.var_identities}
\frac{\delta \gamma(r,X')}{\delta \gamma_{ij}(r,X)} &=\gamma(r,X) \gamma^{ij}(r,X) \delta^{(d-1)}(X-X') \, , \\ \gamma_{ij}(r,X) \, \frac{\delta \sqrt{\gamma(r,X')}}{\delta \gamma_{ij}(r,X)} &= \frac{d-1}{2} \sqrt{\gamma(r,X)} \, \delta^{(d-1)}(X-X') \, , \\
\gamma_{ij}(r,X) \, \frac{\delta R[\gamma(r,X')]}{\delta \gamma_{ij}(r,X)} &= - R[\gamma(r,X)] \delta(X-X') + ... \, , \\
\gamma_{ij}(r,X) \, \frac{\delta \Delta_{\gamma(r,X')}}{\delta \gamma_{ij}(r,X)} &=  - \Delta_{\gamma(r,X)} \delta(X-X') + ... \, ,
\end{align}
where the ellipsis denote boundary terms that we set to zero.

\subsection{Other identities and conventions}
\paragraph{Fourier transforms.}
We define the inverse Fourier transform with a factor of $\frac{1}{2 \pi}$, such that
\begin{equation}\label{eq:FTdelta}
\int_{-\infty}^{\infty} \frac{\dt \omega}{2\pi} e^{-i\omega (t-t')} = \delta(t-t') \,. 
\end{equation}

\paragraph{Coordinates on the sphere.}
The volume form of the unit sphere $\mathds{S}^{d-1}$ is given by 
\begin{align}
    \dt \Omega = \prod_{i=1}^{d-1} \dt \vartheta^i \, \sqrt{\hat{g}}\, , 
\end{align}
where $\vartheta^i$ are the coordinates of the sphere and $\sqrt{\hat{g}}$ is the volume element. Moreover, we define the corresponding delta distribution as follows
\begin{align}
\label{eq:AngularDelta}
    \delta(\Omega- \Omega')  = \frac{1}{\sqrt{\hat{g}}} \prod_{i = 1}^{d-1}\delta(\vartheta^i - \vartheta^{i'})\, . 
\end{align}
Note that functional derivatives for a function of the sphere satisfy
\begin{align}
    \frac{\delta f(\vartheta^i)}{\delta f(\vartheta^{i'})} = \delta(\vartheta^i - \vartheta^{i'})\, . 
\end{align}
The (real) spherical harmonics satisfy the following completeness relation
\begin{equation}
\label{eq:ToolForPropagotors_1}
\sum_{\ell,I} Y_{\ell I}(\Omega) Y_{\ell I}(\Omega') = \delta^{d-1}(\Omega - \Omega')\,.
\end{equation}

\paragraph{Bessel \& Hankel functions.} The integral representation of Dirac delta is given in equation 1.17.13 of \cite {NIST:DLMF}
\begin{equation}\label{eq:ToolForPropagotors_2}
\int_0^\infty \dt \kappa \, \kappa \, J_\nu(\kappa r) J_\nu(\kappa r') = \frac{1}{r} \delta(r - r') \,.
\end{equation}
The integral of a double product is given in equation 10.22.69 of
 \cite{NIST:DLMF} 
\begin{equation}\label{eq:integra_bessel}
\int_{0}^{\infty} \frac{t \dt t}{t^2 - z^2} J_{\nu}(a t) J_{\nu}(b t) = 
\begin{cases}
\frac{1}{2} \pi i J_{\nu}(b z) H_{\nu}^{(1)}(a z) & \text{for } a > b \\
\frac{1}{2} \pi i J_{\nu}(a z) H_{\nu}^{(1)}(b z) & \text{for } b > a
\end{cases}
\quad \text{for } \text{Im}(z) > 0\, . 
\end{equation}
The Wronski determinant of Hankel functions is given in equation 10.5.5 of \cite{NIST:DLMF} 
\begin{align}\label{eq:crossproduct}
   {H^{(2)}_{\nu}}'(\beta r)
    H^{(1)}_{\nu}(\beta r) -{H^{(1)}_{\nu}}'(\beta r)
    H^{(2)}_{\nu}(\beta r)=- \frac{4i}{ \pi \beta r}\, . 
\end{align}

\section{Scalar field in Euclidean space}
\label{sec:EuclideanOnshellAction}
In this appendix, we run the holographic renormalization procedure for the free scalar field in Euclidean space (using $(\tau,r)$-coordinates defined in \eqref{eq:metrictaur}). The benefit of the Euclidean setting is that we have proper divergences without the need of imposing an $i\epsilon$ prescription, in contrast to the Lorentzian version of the computation. In the following, we show how the divergences of the on-shell action are removed by the counterterms. The renormalization of the canonical momentum follows analogously.  

Let us restrict to the massless case $m=0$. The asymptotic solutions of the field equation are similar to \eqref{eq:TRasymptoticFullField}, namely
\small
\begin{align}\label{eq:EucldieanAsymptoticSeries}
    \Phi(r, \tau, \Omega) = r^{-\frac{d-1}{2}}\sum_{\ell, I}Y_{\ell, I}(\Omega)\, \int \frac{\dt\omega}{2 \pi } e^{ -i \omega \tau}\,\left( e^{+|\omega| r}\sum_{k=0}^{\infty}\tilde{\phi}^{(\mathrm{I})}_{k}(\omega, \ell, I)r^{-k} + e^{-|\omega| r}\sum_{k =0}^{\infty}\tilde{\phi}^{(\mathrm{II})}_{k}(\omega, \ell, I) r^{-k}\right)\,. 
\end{align}
\normalsize
The regularised on-shell action \eqref{eq:def:regOS} is
\begin{align}
    S^{\mathrm{reg}}_{\mathrm{os}} = \frac{1}{2}\int_{r = r_0} \dt\tau\, \dt\Omega \, r^{d-1}\, \Phi \partial_r \Phi\, . 
\end{align}
Inserting the expansion \eqref{eq:EucldieanAsymptoticSeries} and dropping the $\ell, \,I$ dependencies of the fields, gives
\small
\begin{align}
\begin{split}
    S_{\mathrm{os}}^{\mathrm{reg}} &= \frac{1}{4\pi}\sum_{\ell, I}\int_{r = r_0} \dt\omega\,\sum_{k, k'=0}^{\infty}\, r^{-k-k'}\left\{ e^{2 |\omega| r}\left(\frac{-\frac{d-1}{2}-k}{r}+|\omega|\right)\tilde{\phi}_{k}^{(\mathrm{I})}( -\omega)\tilde{\phi}^{(\mathrm{I})}_{k'}(\omega)\right. \\
    &\quad\left.+ \left(\frac{-\frac{d-1}{2}-k}{r} + |\omega|\right)\tilde{\phi}_{k'}^{(\mathrm{II})}(-\omega)\tilde{\phi}^{(\mathrm{I})}_{k}(\omega) + \left(\frac{-\frac{d-1}{2}-k}{r}-|\omega|\right)\tilde{\phi}_{k'}^{(\mathrm{I})}(-\omega)\tilde{\phi}_{k}^{(\mathrm{II})}(\omega) \right.\\
    &\quad\left.+ \,e^{-2 |\omega| r}\left(\frac{-\frac{d-1}{2}-k}{r}-|\omega|\right)\tilde{\phi}_{k'}^{(\mathrm{II})}(-\omega)\tilde{\phi}_{k}^{(\mathrm{II})}( \omega)\right\} \\
    & =\sum_{\ell, I}\int_{r = r_0} \dt\omega\, \sum_{k, k'=0}^{\infty}\,  e^{2 |\omega| r }r^{-k-k'}\left(\frac{-\frac{d-1}{2}-k}{r}+|\omega|\right)\tilde{\phi}^{(\mathrm{I})}_{k'}(-\omega)\tilde{\phi}^{(\mathrm{I})}_{k}(\omega) + O\left(r_0^{-1}\right)\, .  
\end{split}
\end{align}
\normalsize
Since $|\omega|\geq 0$, the first part diverges exponentially, while the other parts vanish as $r_0\to\infty$. There are no finite parts in the on-shell action. 
As described in section \ref{sec:Holographicrenormalizationformasslessfields}, the counterterms take the form
\begin{align}
    S^{\mathrm{ct}} =\frac{1}{2} \int_{r = r_0} \dt\tau\, \dt\Omega \, r^{d-1} \,\Phi \frac{\partial_r g}{g} \Phi\, , 
\end{align}
where $g$ is the asymptotic series for the divergent part
\begin{align}
    g = e^{|\omega| r}\sum_{k =0}^{\infty} \tilde{\phi}_{k}^{\mathrm{(I)}}(\omega, \ell, I) r^{-\frac{d-1}{2}-k}\, . 
\end{align}
This branch is going to be the source branch, with the source being  proportional to $\tilde{\phi}_0^{(\mathrm{I})}$. This identification appears to differ from what we did in the Lorentzian case, but this is only the effect of a labeling convention as we have previously fixed the conventions for $\beta_+$ and $\beta_-$.

As we noted in the main text, there is an infinite series of divergent counterterms
\begin{align}
    \begin{split}
    S_{\mathrm{ct}}^{\mathrm{reg}}
    & = \sum_{\ell, I}\int_{r =r_0} \dt\omega \, \sum_{k, k' =0}^{\infty}\, \left\{ e^{2 |\omega| r}\left(|\omega| - \frac{\frac{d-1}{2} +k}{r}\right)r^{-k-k'}\tilde{\phi}^{(\mathrm{I})}_{k'}(-\omega, \ell, I)\tilde{\phi}^{(\mathrm{I})}_{k}(\omega, \ell, I) \right.\\ 
    &\qquad \left. +  |\omega| \left( \tilde{\phi}^{(\mathrm{II})}_{0}(-\omega,\ell, I)\tilde{\phi}^{(\mathrm{I})}_{0}(\omega, \ell, I)+\tilde{\phi}^{(\mathrm{I})}_{0}( -\omega, \ell, I)\tilde{\phi}^{(\mathrm{II})}_{0}(\omega, \ell, I)\right)\right\} + O(r_0^{-1})\, , 
    \end{split}
\end{align}
which exactly cancel the divergences in the on-shell action and leave us with a non-zero finite renormalized on-shell action
\begin{align}
\label{eq:TauR:RenOnshellAction}
    S^{\mathrm{ren}}_{\mathrm{os}} = -\frac{1}{2 \pi} \sum_{\ell, I}\int \dt \omega \,|\omega|\, \tilde{\phi}^{(\mathrm{II})}_{0}(-\omega, \ell, I)\,\tilde{\phi}_{0}^{(\mathrm{I})}( \omega, \ell, I)\, . 
\end{align}

\section{Factorization of the equations of motion}
\label{sec:factorization}
In this appendix, we want to present a different perspective on the function $f$ that appears in the counterterm action \eqref{eq:AnsatzCounterterms} for the free scalar field. For holographic renormalization in AdS/CFT, this perspective has been introduced in \cite{Papadimitriou:2003is}.

The equations of motion of the free scalar field in pure Minkowski space time (but also AdS space times) in $(t,r)\, , (v,r)$ and $(\tau, r)$-coordinates are of the form
\begin{align}
\label{eq:generalizedWaveEquation}
    \left(\partial_r^2 + A(r,X,  \partial_X)\partial_r + B(r,X, \partial_X)\right)\Phi =0, 
\end{align}
where $A(r,X,\partial_X)$ and $B(r,X,\partial_X)$ are functions of $r$ (``the holographic direction") and, generically, differential operators $\partial_X$ of the transverse directions\footnote{We use the term ``transverse directions" not so precisely to indicate all the remaining dependence excluding $r$. } $X$. These functions $A$ and $B$ are easily read-off from \eqref{eq.eom_tr}, \eqref{eq.eom_vr} and \eqref{eq.eomtaur}, namely
\begin{subequations}
\begin{align}
&(\tau,r):\quad A=\frac{d-1}{r},\quad B= \frac{1}{r^2}\Delta_{\mathds{S}^{d-1}}  - m^2 + \partial_{\tau}^{2}\,,\\
&(t,r):\quad A=\frac{d-1}{r},\quad B=\frac{1}{r^2}\Delta_{\mathds{S}^{d-1}}  - m^2 - \partial_{t}^{2}\,,\label{eq:AB-TR}\\
&(v,r):\quad A= \frac{d-1}{r} + 2 \partial_v\,,\quad B=\frac{1}{r^2}\Delta_{\mathds{S}^{d-1}} - \left(m^2 - \frac{d-1}{r}\partial_v\right)\,.
\end{align}
\end{subequations}
Let us denote two generic linearly independent asymptotic solutions
as $g_-$ and $g_+$ and superimpose them with $r$-independent coefficients $\phi_s(X)$ and $\phi_v(X)$ as
\begin{align}
\label{eq:generalBranchSplit}
    \Phi(r,X) = g_{-}(r,X, \partial_X)\phi_s + g_{+}(r,X, \partial_X) \phi_v. 
\end{align}
Note that $\phi_s$ and $\phi_v$ do depend on the transverse coordinates and are obviously related to $\phi^{(\mathrm{I})}$ and $\phi^{(\mathrm{II})}$ appearing in section \ref{sec:FreeScalarFieldSolutions}. 

Furthermore, the equation \eqref{eq:generalizedWaveEquation} can be factorized in the form  
\begin{align}\label{eq:factorisedeq}
    (\partial_r  - f_+)\,(\partial_r - f_-)\,\Phi = 0, 
\end{align}
where by comparison with \eqref{eq:generalizedWaveEquation},
\begin{align}
\label{eq:definition_fFACTOR}
    f_+f_-- \partial_r f_- -B =0, \qquad f_++f_-=-A.
\end{align}
Combining both equations in
\eqref{eq:definition_fFACTOR} we obtain a non-linear first order differential equation of Riccati type for $f_-$
\begin{align}
\label{eq:Factorization}
    (-A - f_-)f_- - \partial_r f_- -B =0\,. 
\end{align}
This equation is exactly the Riccati equation \eqref{eq:RiccatiEquation} in $(t, r)$-coordinates. 
The solution of \eqref{eq:Factorization} may be expressed as 
\begin{align}
\label{eq:generalshapef}
f_- = (\partial_r g_{-})\,g_{-}^{-1}\,,
\end{align}
where the differential operator $g_-$ satisfies \eqref{eq:generalizedWaveEquation} with $\Phi$ substituted for $g_-$. Note that, by construction, $g_-$ satisfies the following differential equation
\begin{align}
\label{eq:projectingpropertyFactorization}
    (\partial_r - f_-) \,g_-(r,X) = 0 \, .
\end{align}
Similarly, we may characterise $g_+$ as the solution of the equation 
\begin{align}
    (\partial_r - f_+)\,(\partial_r - f_-)\,g_{+}(r,X) & = 0\,, \quad \text{with} \quad 
    (\partial_r - f_-)\,g_{+}(r,X) \neq 0\,.
\end{align}
The significance of this observation lies in the fact that the kernel of the differential operator $(\partial_r-f_-)$ consists of the $g_-$ branch of the two linearly independent solutions. 
The reader familiar with AdS/CFT or with section \ref{sec:reviewAdS}, would have already noted that this is exactly what is required to holographically renormalise the AdS scalar canonical momentum \eqref{eq:AdScanonicalMomentum} (or action), after $g_-$ is identified with the non-normalisable branch and $g_+$ with the normalisable brach.
 In this respect, the flat case is remarkably similar.  In fact, $(\partial_r -f_-) \Phi$ corresponds to the renormalised momentum  discussed in section \ref{sec:holoren_tr} and the property above ensures that the canonical momentum has roughly the same asymptotic behaviour as $g_+$. 

\section{Details on the calculations for interactions}
\label{sec:InteractionDetails}
\subsection{Derivation of the bulk-to-bulk propagator}
\label{sec:bulkbulkDETAILS}
In this appendix we determine the bulk-to-bulk propagator as the solution of 
\begin{equation}
\label{eq:defBulkBulk}
(\Box - m^2) \,G(x; x') = \frac{1}{\sqrt{|g|}} \delta^{d+1}(x - x')\, , 
\end{equation}
with appropriate boundary conditions corresponding to Feynman propagators along the lines of \cite{Moreira:1995he}.
The delta function in \eqref{eq:defBulkBulk} is defined as 
\begin{equation}
\label{eq:defBulkDelta}
\frac{1}{\sqrt{|g|}} \delta^{d+1}(x - x')=\frac{1}{r^{d-1}} \delta(r - r')\delta^{d-1}(\Omega - \Omega')\delta(t- t')\, , 
\end{equation}
with the angular part defined in \eqref{eq:AngularDelta}. The associated
spectral problem,
\begin{equation}
(\Box - m^2) \psi = \mu \psi \quad \Longleftrightarrow \quad (\Box - m^2 - \mu) \psi = 0\, , 
\end{equation}
with $\psi$ being regular in the interior, has the regular solution
\begin{equation}
\label{eq:BULKBULKparticularsolution}
\psi(x) = \frac{1}{r^{\frac{d-2}{2}}} \frac{e^{-i\omega t}}{\sqrt{2\pi}} J_\nu \left( \kappa \, r \right) Y_{\ell I}(\Omega)\,,
\end{equation}
where we defined
\begin{equation}
\kappa := \sqrt{\omega^2+i\epsilon - m^2 - \mu} \,\in \mathbb{R}_+ \, . 
\end{equation}
The $ i\epsilon $ prescription is chosen according to \eqref{eq:iepsilonprescription}.\footnote{For other propagators (such as retarded or advanced), the $ i\epsilon $ prescriptions has to be modified. In particular, for retarded propagators we have to use $ (\omega + i\epsilon)^2 \sim \omega^2 + 2i\epsilon \omega$ instead of $ \omega^2 + i\epsilon $.} Using the identities \eqref{eq:ToolForPropagotors_1} and  \eqref{eq:ToolForPropagotors_2}
we find that the solution \eqref{eq:BULKBULKparticularsolution} obeys the following completeness relation
\begin{equation}
\sum_{\ell, I} \int_{-\infty}^{\infty} d\omega \int_{0}^{\infty} \dt \kappa \, \kappa \, \psi_{\omega \ell I \kappa}(x) \, \psi_{-\omega \ell I \kappa}(x') =\frac{1}{\sqrt{-g}} \delta^{d+1}(x - x').
\end{equation}
This enables us to write the bulk-to-bulk propagator as
\begin{equation}
\label{eq:rawbulkbulk}
G(x; x') = \sum_{\ell, I} \int_{-\infty}^{\infty} d\omega \int_{0}^{\infty} d\kappa  \frac{\kappa}{\omega^2 - \kappa^2 - m^2 + i\epsilon} \psi_{\omega \ell I \kappa}(x)  \psi_{-\omega \ell I  \kappa}(x')\, , 
\end{equation}
which by construction satisfies \eqref{eq:defBulkBulk}. 
We can integrate over $ \kappa \in [0, \infty) $ using the identity \eqref{eq:integra_bessel} in the form
\begin{align}
 -\int_{0}^{\infty} \dt \kappa \, \frac{\kappa}{\kappa^2 - \kappa_0^2} J_{\nu}(k r) J_{\nu}(k r')
= 
\begin{cases}
-\frac{1}{2} \pi i J_{\nu}(\kappa_0 r) H_{\nu}^{(1)}(\kappa_0 r') & \text{for } r' > r \\
-\frac{1}{2} \pi i J_{\nu}(\kappa_0 r') H_{\nu}^{(1)}(\kappa_0 r) & \text{for } r > r'
\end{cases}\,\,,
\end{align}
where we defined
\begin{equation}
\kappa_0^2 := \omega^2 - m^2 + i\epsilon\, . 
\end{equation}
The bulk-to-bulk propagator $G(x; x')$ is thus given by
\begin{align}
\begin{split}
G(x; x') = & -\frac{i\pi}{2}\sum_{\ell, I} \int_{-\infty}^{\infty} \frac{\dt \omega}{2\pi} e^{-i\omega(t - t')} \frac{1}{(r r')^{\frac{d-2}{2}}} Y_{\ell I}(\Omega) Y_{\ell I}(\Omega')\\& \times
\begin{cases}
J_{\nu} \left( \sqrt{\omega^2 - m^2} \, r' \right) H_{\nu}^{(1)} \left( \sqrt{\omega^2 - m^2} \, r \right) & \text{for } r > r' \\
J_{\nu} \left( \sqrt{\omega^2 - m^2} \, r \right) H_{\nu}^{(1)} \left( \sqrt{\omega^2 - m^2} \, r' \right) & \text{for } r' > r
\end{cases} \, , 
\end{split}
\end{align}
with $ \nu = \ell + \frac{d-2}{2} $ and we suppressed the $i\epsilon$ terms due to displaying purposes. 
When promoting $\nu$ again to the differential operator \eqref{eq:nuDef}, we can perform the sum using \eqref{eq:ToolForPropagotors_1} to obtain the bulk-to-bulk propagator
\begin{align}
\begin{split}
G(x, x') = & -\frac{i\pi}{2 }\int_{-\infty}^{\infty}  \frac{\dt \omega}{2\pi} e^{-i\omega(t - t')} \frac{1}{(r r')^{\frac{d-2}{2}}} \\& \times
\begin{cases}
J_{\nu} \left( \sqrt{\omega^2 - m^2} \, r' \right) H_{\nu}^{(1)} \left( \sqrt{\omega^2 - m^2} \, r \right)\delta(\Omega - \Omega') & \text{for } r > r' \\
J_{\nu} \left( \sqrt{\omega^2 - m^2} \, r \right) H_{\nu}^{(1)} \left( \sqrt{\omega^2 - m^2} \, r' \right)\delta(\Omega - \Omega') & \text{for } r' > r
\end{cases} \, .
\end{split}
\end{align}

\subsection{Three-point function in frequency space}
\label{sec:3ptFunctionFrequencyCalcs}

We would like to work out \eqref{eq:3ptFreqStartingpoint}. 
It appears to be amenable to compute the three-point function in frequency domain (we do not apply the decomposition in spherical harmonics \eqref{eq:conventionsphericalharmonics} here)
\begin{align}
    F(\omega_j, \Omega_j)\, \equiv \langle \mathcal{O}(\omega_1, \Omega_1)\mathcal{O}(\omega_2, \Omega_2)\mathcal{O}(\omega_3, \Omega_3) \rangle_{\mathrm{tree}}\, . 
\end{align}
We start from \eqref{eq:3ptFreqStartingpoint} and keep in mind that the bulk-to-boundary propagators are defined in \eqref{eq:defKs} and differ from the one defined in \eqref{eq:bulkboundaryPropagator} by a factor of $\mathrm{sign}(\omega)^{\ell}$. 
\small
\begin{align}
\begin{split}
     F\left(\omega_j, \Omega_j\right)=&\int \dt r \int \dt^d X \sqrt{\gamma} e^{-it(\omega_1 + \omega_2 + \omega_3)} \\
    & \times \prod_{j =1 }^3 \left(\mathrm{sign}(\omega_j)^{\ell_j}\, r^{-\frac{d-2}{2}} \sqrt{2 \pi} e^{-i\pi \frac{2 \nu_j +1}{4}}J_{\nu_j}(|\omega_j|r)|\omega_j|^{\frac{1}{2}}\delta\left(\Omega - \Omega_j\right)\right)\\
    \end{split}\\
    \begin{split}
    \phantom{F\left(\omega_j, \Omega_j\right)}
    =&\int \dt r \int \dt \Omega \,r^{d-1}\, \delta\left(\sum_{j=1}^3\omega_j\right) \prod^3_{j = 1}\left(\left(\frac{|\omega_j|}{2 \pi}\right)^{\frac{d-1}{2}}e^{\frac{-i\pi(d-1)}{4}}\sum_{\ell_j, I_j} \right)\\
    & \times \prod_{j =1 }^3 \left(\mathrm{sign}(\omega_j)^{\ell_j}(r|\omega_j|)^{-\frac{d-2}{2}} (2 \pi)^\frac{d}{2} (-i)^{\ell_j}J_{\nu_j}(|\omega_j|r)Y_{\ell_j, I_j}(\Omega)Y_{\ell_j, I_j}(\Omega_j)\right)\\
    \end{split}\\
    \begin{split}
    &\phantom{F\left(\omega_j, \Omega_j\right)}= \int \dt^d  \vec{x}\, \prod^3_{j = 1}\left(\left(\frac{|\omega_j|}{2 \pi}\right)^{\frac{d-1}{2}}e^{\frac{-i\pi(d-1)}{4}}\right)\delta(\omega_1 + \omega_2 + \omega_3) \\
    & \phantom{F\left(\omega_j, \Omega_j\right) -} \times e^{-i\left( \vec{x}\cdot \left(\mathrm{sign}(\omega_1)\hat{x}_1|\omega_1|+\mathrm{sign}(\omega_2)\hat{x}_2|\omega_2|+\mathrm{sign}(\omega_3)\hat{x}_3|\omega_3 |\right) \right)}  \, ,
    \end{split}
\end{align}
\normalsize
where we applied \eqref{eq:Bessel-sphericalharmonicsddim} and the fact that $(-1)^{\ell}$ acting on the spherical harmonics is the antipodal map on the angles. Recall that $\Omega$ here just denotes all the angles, which are expressed as a unit vector in order to make contact with \eqref{eq:Bessel-sphericalharmonicsddim}.
In this form, we can perform the integral over the space point to obtain
\begin{align}
\label{eq:3pFunctionIntermediate}
\begin{split}
    \langle \mathcal{O}(\omega_1, \Omega_1)\mathcal{O}(\omega_2, \Omega_2)\mathcal{O}(\omega_3, \Omega_3) \rangle_{\mathrm{tree}} =& \prod^3_{j = 1}\left(\left(\frac{|\omega_j|}{2 \pi}\right)^{\frac{d-1}{2}}e^{\frac{-i\pi(d-1)}{4}}\right)\\
    & \times \delta \left(\hat{x}_1\omega_1+\hat{x}_2\omega_2+\hat{x}_3\omega_3 \right)\delta(\omega_1 + \omega_2 + \omega_3)\, , 
    \end{split}
\end{align}
which corresponds to \eqref{eq:3pFunctionIntermediatemain} in the main text.

\bibliography{biblio}
\bibliographystyle{JHEP}

\end{document}